\documentclass[10pt,journal,compsoc]{IEEEtran}

\usepackage[nocompress]{cite}

\usepackage{amsmath}
\usepackage[pdftex]{graphicx}
\usepackage{balance}       
\usepackage[pdflang={en-US},pdftex]{hyperref}
\usepackage{xcolor}

\usepackage{multirow}
\usepackage{subfigure}
\usepackage[noend]{algorithmic}
\usepackage[ruled]{algorithm}
\usepackage{textcomp}
\usepackage[normalem]{ulem} 
\usepackage{soul} 

\newenvironment{myalgorithm}[2][\columnwidth]
{ \begin{center}
   \begin{minipage}{#1}
     \begin{algorithm}[H]
     \caption{#2}
     \algsetup{linenosize=\small, linenodelimiter=.}
       \begin{algorithmic}[1]}
{      \end{algorithmic}
     \end{algorithm}
   \end{minipage}
 \end{center}}
 
\usepackage{tikz}

\newcommand*\circled[1]{\tikz[baseline=(char.base)]{
		\node[shape=circle,draw,inner sep=1pt] (char) {#1};}}

\newcommand{\sdk}{AFV}

\begin{document}
%
\title{Seamless 
Resources Sharing in Wearable Networks by Application Function Virtualization}

%

\author{Harini Kolamunna, Kanchana Thilakarathna, Diego Perino, Dwight Makaroff and Aruna Seneviratne
\IEEEcompsocitemizethanks{\IEEEcompsocthanksitem H. Kolamunna, K. Thilakarathna, and A. Seneviratne are with the School of EE\&T, University of New South Wales and Data61-CSIRO, Eveleigh, NSW 2015, Australia. E-mail: \{harini.kolamunna, kanchana.thilakarathna, aruna.seneviratne\}@data61.csiro.au\protect\\

\IEEEcompsocthanksitem D. Perino is with Telefonica Research. Email: {diego.perino@telefonica.com}\protect\\

\IEEEcompsocthanksitem D. Makaroff is with University of Saskatchewan. E-mail: {makaroff@cs.usask.ca}}}


\IEEEtitleabstractindextext{%
\begin{abstract}
The prevalence of smart wearable devices is increasing exponentially and we are witnessing a wide variety of fascinating new services that leverage the capabilities of these wearables. Wearables are truly changing the way mobile computing is deployed and mobile applications are being developed. It is possible to leverage the capabilities such as connectivity, processing, and sensing of wearable devices in an adaptive manner for efficient resource usage and information accuracy within the personal area network.  We show that application developers are not yet taking advantage of these cross-device capabilities, however, instead using wearables as passive sensors or simple end displays to provide notifications to the user. We thus design \sdk~(Application Function Virtualization), an architecture enabling automated dynamic function virtualization and scheduling across devices in a personal area network, simplifying the development of the apps that are adaptive to context changes. \sdk~provides a simple set of APIs hiding complex architectural tasks from app developers whilst continuously monitoring the user, device and network context, to enable the adaptive invocation of functions across devices. We show the feasibility of our design by implementing \sdk~on Android, and the benefits for the user in terms of resource efficiency, especially in saving energy consumption, and quality of experience with multiple use cases.
\end{abstract}

}

\maketitle
\IEEEdisplaynontitleabstractindextext
\IEEEpeerreviewmaketitle

\IEEEraisesectionheading{\section{Introduction}\label{sec:intro}}

\IEEEPARstart{S}{mart} wearable devices such as smartphones, tablets, smartwatches and fitness bands enable mobile users 
to form Personal Area Networks (PANs).
%
Some of the devices in the PAN are capable of providing the same functionality.
For instance, a fitness band, a smartwatch and a smartphone are each likely to have an accelerometer, a gyroscope, and a heart rate monitor. Similarly, a number of devices on the PAN may have  direct Internet connectivity, providing multiple network interfaces. Furthermore, some wearable devices 
will have sufficient computing resources to perform functions such as data encoding, compression and encryption, while others may not. 

Previous analysis  of several popular wearable health and fitness tracking applications \cite{Kolamunna2016} shows that app developers tend not to leverage 
available resources on other devices.
For examples, the smartwatch pedometer will still use its own accelerometer when the battery level is low, despite an accelerometer being available on a fully-charged smartphone. 
The primary reason for relying on local resources only is the app developers reliance on the APIs provided by the target device.\footnote{These APIs abstract a large spectrum of functions (e.g., sensing, communication) that are actually implemented by the operating systems (OS) or third party libraries and executed in the device where the application is running.} 

To harness the collective capabilities of PAN devices, developers have to implement each  app individually to utilize the distributed device resources, managing the cost of running these functions in each device and communication costs explicitly. 
This requires developers to have a wider understanding of distributed systems and increases the complexity of mobile app development. 
Therefore, an architecture that takes user and device context into account, enabling app developers to utilize all PAN resources easily is needed.



In this paper, we present such an architecture that extends the concept of network function virtualization \cite{MijumbiComSurTot15} to device functions. We show the proposed architecture's advantages to users and application developers and make the following contributions:
\begin{itemize}
\item Provide an architecture (\sdk) that enables wearable/mobile application function virtualization for the development of adaptive wearable/mobile applications.
\item Design \sdk's~inter-device and intra-device communication protocols to minimize the overhead added by the architecture and maximizes the advantages.

\item Propose and evaluate a greedy heuristic algorithm for adaptive function allocation across devices considering the available resources and dynamic context of devices in the PAN. 


\item Demonstrate  \sdk's ability to adapt to context changes dynamically and demonstrate user and performance benefits of using \sdk~using a number of applications, 
and their usage.

\end{itemize}

The rest of this paper is organized as follows. 
We first present the \sdk~architecture, describing each module of \sdk~followed by the context-aware adaptive function allocation algorithm. Next, we show the realization of \sdk~with implementation in Android and the experimental calibrations. Performance evaluation with simulation and experiments are presented in details with the experimented use cases of \sdk. Finally, we overview related work and complementary systems, and provide conclusions with future work.

\section{\sdk: Application Function Virtualization}
\label{sec:framework}

\begin{figure}[ht]
\vspace {-3mm}
\centering 
\includegraphics[width=0.47\textwidth]{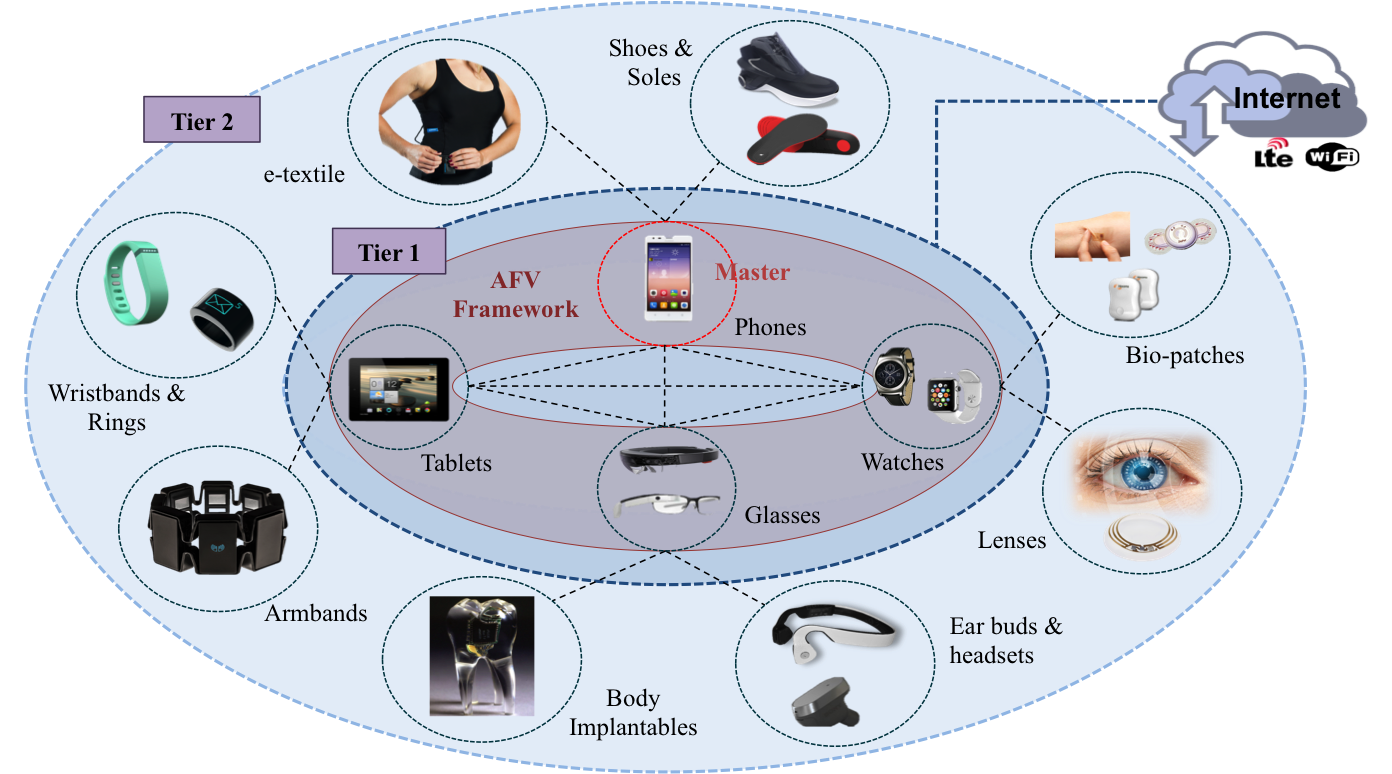}
\vspace {-2mm}
\caption{An overview of an personal wearable network.}
\label{fig:pwn}
\end{figure}

Current PAN devices can be divided into two broad categories, which we refer to as Tier 1 and Tier 2, depicted in Figure~\ref{fig:pwn}. Tier 1 devices, (e.g. smartglasses, smartwatches, smartphones), are relatively more resourceful than Tier 2 devices. Tier 2 devices, (e.g. smartshirts and bio-patches) simply carry out sensing functions. In contrast to Tier 2 devices, Tier 1 devices have the following additional features: 
a) availability of heterogeneous long-range network \textit{connectivity} (WiFi, cellular), and b) ability to \textit{process} sensed and received information. 

Therefore, Tier 1 devices may be equipped with a rich-set of sensors, multiple connectivity interfaces, storage and computing power to perform wide set of functions such as compression, encoding, rendering, intrusion detection, firewall filtering and encryption. Some of these resources are context dependent, e.g. WiFi connectivity will only be available in limited locations, GPS location will not be available indoors and devices may be disconnected if the battery is depleted. 

Since, Tier 2 devices provide complementary and specific functionality that may be duplicated on available Tier 1 devices, an application does not usually need sensors on all devices to be active simultaneously. It is also the case that different sensors have differing output quality and resource requirements that make the selection of functionality for optimal user benefit at run-time a challenge.
Thus if all the resources available on a PAN can be effectively utilized in an automated fashion, depending on the user/device context, it should be possible to significantly improve both the dimensions of user utility: functional requirements (precision, accuracy), and performance (energy consumption, latency). 

As mentioned, application developers do not effectively utilize all the available resources in a PAN, when developing applications. We believe to facilitate the use of PAN wide resources, that it is necessary to provide methods for 
designers to seamlessly access the resources, taking in to account the context of use. 
We further believe this is achievable by virtualizing the commonly used functions distributed across multiple devices on a PAN, and orchestrating the use of the functions depending the context of use, and the system state, which we refer to as \sdk. In so doing, it is possible to accomplish the goals of a) minimizing the development effort for application developers; b) minimizing the configuration burden of the users;  and c) maximizing the user quality of experience.


\begin{figure}[t]
\centering 
\includegraphics[width=0.47\textwidth]{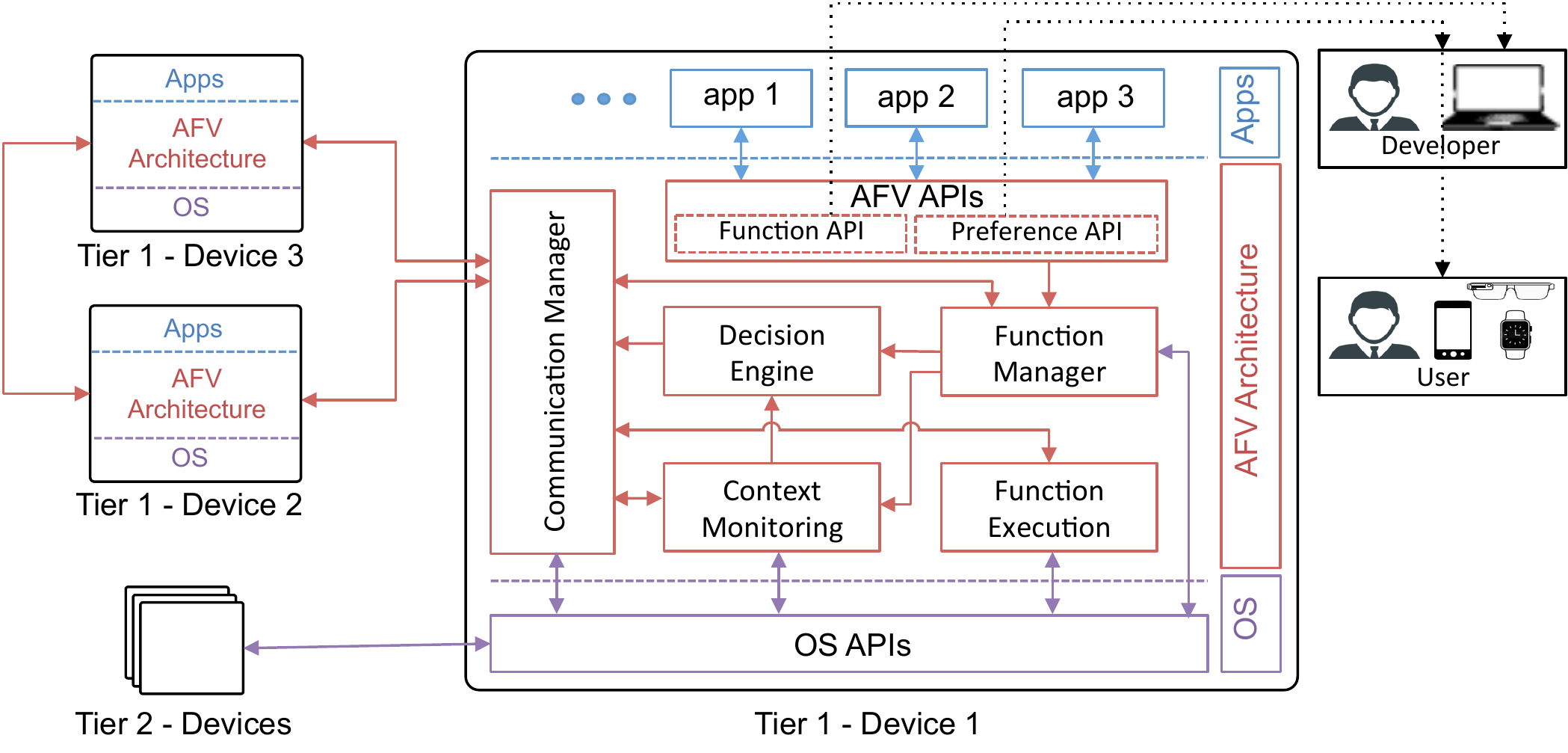}
\caption{Overview of the \sdk~and logical connectivity among devices.}
\label{fig:framework}
\vspace{-3mm}
\end{figure}

\subsection{\sdk~Architecture}

\sdk~is a collaborative platform, which runs on the Tier 1 devices and interacts with the Tier 2 devices in a PAN as shown in Figure~\ref{fig:pwn}, makes  available all potential resources to application developers and/or users seamlessly through APIs. 

Figure \ref{fig:framework} provides a schematic view of the \sdk~architecture. 
To use \sdk, applications register requests for the required functions via the \emph{\sdk~APIs} explained in Section \ref{subsec:apis}. Within \sdk, application function registration requests are handled by the \emph{Function Manager} module as described in Section \ref{subsec:fmanager}. The \emph{Context Monitoring} module periodically monitors device and user context as detailed in Section \ref{subsec:context}. One of the Tier 1 devices on a PAN is elected as the \emph{Master Device}. 
This can either be done by the user or automatically based on a set of criteria such as the lowest ratio between current state of charge and energy usage (i.e., the Tier 1 device that would be least affected by the master tasks).

\begin{figure*}[tb!]
	\centering 
	\includegraphics[width=1\textwidth]{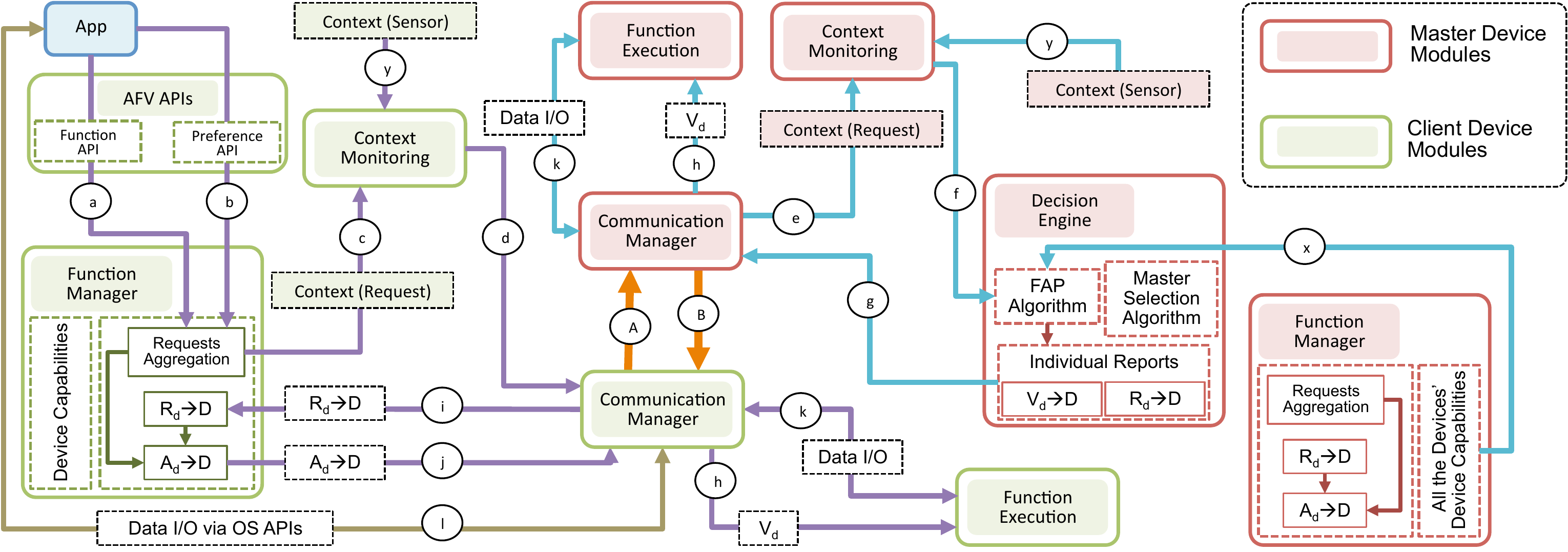}
	\vspace {-3mm}
	\caption{Diagrammatic representation of \sdk.}
	\vspace {-3mm}
	
	\label{fig:sch}
\end{figure*}

The automated \emph{Master Device} selection is done in the \emph{Decision Engine} (described in Section \ref{sec:DecisionEngine}) as a collective process of all the Tier 1 devices. 
The \emph{Decision Engine} of the selected \emph{Master Device} then determines the optimal mapping of each application function request to a function provided by a device on the PAN. All the other Tier 1 devices transfer context changes to the \emph{Master Device} to be used by the \emph{Decision Engine}. The \emph{Communication Manager} maintains efficient Inter-device and Intra-device communication 
as detailed in Section \ref{subsec:cmanager}. Finally, the \emph{Function Execution} module performs function invocation on devices as described in Section \ref{subsec:fexec}. Figure \ref{fig:sch} shows the diagrammatic representation of the flow of events in AFV when a request is made, which is explained throughout this section.  

\subsection{\sdk~APIs}
\label{subsec:apis}

\sdk~provides two main types of APIs: \emph{Function APIs} that are executed during run time and \emph{Preference APIs} that are executed at  application start-up.

\subsubsection{Function APIs}
For each supported function, \sdk~provides a specific API to the developer. As such, the \emph{\sdk~APIs} augments the existing APIs provided by the operating system. 

As an example, Figure~\ref{fig:apiexample} describes how the \emph{\sdk~API} for the sensing function, enhances the API provided by Android. 
The \texttt{onSensorChanged} and  \texttt{unregisterListener} Android APIs are not changed, but are simply reimplemented as  \texttt{AFVonSensorChanged} and  \texttt{AFVunregisterListener} to be consistent with other APIs. 

The \texttt{registerListener} API requires simple modifications. 
\texttt{AFVregisterListener} augments \texttt{maxReportLatencyUs} input to define the data exchange frequency between applications and devices, and the \texttt{precision} input indicate the required measurement accuracy with respect to absolute correctness, provided as a list of contexts and ranges 
(e.g., $<$running, $<$5\%, 10\%$>$$>$, $<$walking, $<$5\%, 15\%$>$$>$). In addition, the optional \texttt{mapping} parameter which provides a list of (context, device) pairs (e.g., $<$walking, smartphone$>$, $<$sitting, smartwatch$>$), is used to enable the developer to force the \textit{Decision Engine} to select a particular device where possible. 

All \emph{Function APIs} are designed with a similar structure with minimal changes to the existing APIs to reduce the complexity of \sdk~for the application developers. 

\subsubsection{Preference APIs}
There may be different preferred configurations for the user, application and the device itself when executing an application. These configurations are managed in \sdk~via the \emph{Preference APIs}. \sdk~provides three types of \emph{Preference APIs}, namely application (\texttt{setAppPrefs()}), user (\texttt{setUserPrefs()}) and device (\texttt{setDevicePrefs()}), which have the same structure. An example of \texttt{setAppPrefs()} is illustrated in Figure~\ref{fig:preferenceapi}. 
The configurations (application, user and device) could be conflicting. This is mitigated by certain possible configurations being mandated, or eliminated. 
In the current implementation the following order of priority is used: Device, User, Application. 

Consider a scenario where the developer (via \texttt{setAppPrefs()}) may specify a fitness tracking application to synchronize data with an Internet server, whenever data connectivity is available. However, the user may wish to override the application's settings (via \texttt{setUserPrefs()}) by configuring the application to synchronize data only when WiFi connectivity is available. 
%
For this example, the \texttt{preference} input in \texttt{setAppPrefs()} is $<$connectivity, 0$>$, where ``0" denotes ``any network". And the \texttt{preference} input in \texttt{setUserPrefs()} is $<$connectivity, 1$>$, where ``1" denotes ``WiFi network".
However, the user \emph{Preference API} is not directly exposed to the user; rather, it is leveraged by the developer to take user preferences as inputs. For instance, the UI of the application could provide an interface with radio buttons or drop-down lists to allow the user to select required or forbidden device/context mappings.

Additionally, each device may have configurations defined by the manufacturer. For example, iOS devices are recommended to operate below 35 \textdegree C.\footnote{\url{https://support.apple.com/en-au/HT201678}} These configurations are set by the developer via the \texttt{setDevicePrefs()}, so any user specified configurations beyond this is ignored.

All the preferences received via \emph{Preference APIs} and functional requests received via \emph{Function APIs} are first transferred to the \emph{Function Manager} for the management (cf. \circled{a} and \circled{b} in Figure \ref{fig:sch}). 

\begin{figure}[tb]
\small

\textbf{Android}

\texttt{abstract void onSensorChanged(SensorEventÐ~event)}

\texttt{boolean registerListener(SensorEventListener $~~~~~~~$ listener, Sensor sensor, int sampling $~~~~~~~$ PeriodUs, int maxReportLatencyUs)}\\

\texttt{void unregisterListener(SensorEventListener $~~~~~~~$ listener,   Sensor sensor)}\\

\textbf{\sdk~}

\texttt{abstract void \sdk onSensorChanged(\sdk SensorEvent $~~~~~~~$ event)}

\texttt{boolean \sdk registerListener $~~~~~~~$(\sdk SensorEventListener~listener, 
 $~~~~~~~$ \sdk Sensor sensor, int samplingPeriodUs, $~~~~~~~$ int maxReportLatencyUs, List<Context, $~~~~~~~$  <int,int> > precision, List<Context, $~~~~~~~$ Device> mapping)}\\

\texttt{void \sdk unregisterListener(\sdk SensorEventListener $~~~~~~~$ listener, \sdk Sensor sensor)}\\

\caption{Example of Function APIs provided by \sdk~and by Android.}
\label{fig:apiexample}
\vspace{-2mm}

\end{figure}

\begin{figure}[tb]
\small
\textbf{\sdk~}\\

\texttt{void setAppPrefs(\sdk Application appName,  $~~~~~~~$   \sdk Device deviceName, List<Context, $~~~~~~~$ int> preference)}
\caption{Example of Preference APIs provided by \sdk.}
\label{fig:preferenceapi}
\vspace{-5mm}

\end{figure}

\subsection{Function manager}
\label{subsec:fmanager}
The \emph{Function Manager} is responsible for 
(i) keeping track of device capabilities in terms of supported functions and their associated costs, and (ii) managing the function registration requests and preferences received from \sdk-enabled applications. 
The \emph{Function Manager} in the \emph{Master Device} additionally stores the supported functions and associated costs for all Tier 1 devices and their paired (passive) Tier 2 devices in the PAN. These stored information is transferred to the \emph{Decision Engine} (cf. \circled{x} in Figure \ref{fig:sch}).

A pre-defined list of supported functions of each device is provided with \sdk~architecture. 
The associated cost of each function is composed of two main components; the cost of executing the function and the cost of exchanging inputs/outputs between the requested application and the function running device. Costs would be energy, monetary, latency or any other form. The associated costs are either be provided in the list, which is provided with \sdk~architecture, or obtained at the initialization using methods available in the devices (e.g., \emph{getPower()} method in Android). 
Each time a new \sdk-enabled Tier 1 device or a passive Tier 2 device joins the PAN, the \emph{Function Manager} of the Tier 1 device first discovers the supported functions, and then it announces the availability of functions and related costs to the \emph{Master Device} (cf. Section~\ref{subsec:cmanager}). 


Next key role of the \emph{Function Manager} is the management of function requests and preferences. The preferences received via \emph{Preference APIs} are considered for the requests where applicable, and apply bound conditions to the requests. For an instant, if the user preferred to use WiFi connectivity, the Internet connectivity function requests from that particular application is bound with the condition to use WiFi connectivity only. 
The \emph{Function Manager} aggregates multiple registration requests for the same function from different \sdk-enabled applications, and only invokes the function on one of the devices where possible by considering the preferences (cf. Requests Aggregation in Figure \ref{fig:sch}). 
Any change in the list of registration requests ($R_d$) 
notifies as a change of context and transferred to the \emph{Context Monitoring} module of the device (cf. \circled{c} in Figure \ref{fig:sch}). Eventually, this is transferred to the \emph{Context Monitoring} module of the \emph{Master Device}  (cf. \circled{d},\circled{A},\circled{e} in Figure \ref{fig:sch}), and then triggers the \emph{Decision Engine} of the \emph{Master Device} (cf. \circled{f} in Figure \ref{fig:sch}). 

After each re-evaluation of the function allocation in the \emph{Decision Engine} of the \emph{Master Device}, the \emph{Function Manager} receives the mapping between function registration requests $r \in R_d$ and the device $d \in D$ selected to execute the function ($R_d \mapsto D$) (cf. \circled{g},\circled{B},\circled{i} in Figure \ref{fig:sch}). This is re-mapped to each function requesting \sdk-enabled application $a \in A_d$ and the device $d \in D$ selected to execute the function ($A_d \mapsto D$). Finally, the \emph{Function Manager} transfers the $A_d \mapsto D$ mapping to the \emph{Communication Manger} in order to perform the data transfer from/to the \sdk-enabled applications (cf. \circled{j} in Figure \ref{fig:sch}).
%
%


\begin{table}[ht]
\caption{Example set of context information and mapped system objectives.}
\centering
\scriptsize
\begin{tabular}{|l l l|}
\hline

\multicolumn{2}{|l|}{\textbf{System objective}} & \textbf{Context} \\\hline
\hline
\multicolumn{3}{|l|}{\bf (i) Maximizing the Functional Quality}\\\hline
\multicolumn{2}{|l|}{Precision of fitness tracking} & Moving status (i.e. walking or not)\\\hline
\multicolumn{2}{|l|}{Network throughput} & Average link speed of the network\\\hline\hline
\multicolumn{3}{|l|}{\bf (ii) Energy Utilization}\\\hline
\multicolumn{2}{|l|}{Extend the battery uptime} & Battery level, Default energy usage\\\hline\hline
\multicolumn{3}{|l|}{\bf (iii) Minimizing the Monetrary Cost}\\\hline
\multicolumn{2}{|l|}{Not exceeding the cap data level} & Each device's connected network\\
\hline

\end{tabular}
\label{tab:context}
\vspace{-5mm}
\end{table}

\subsection{Context monitoring}
\label{subsec:context}
We consider the context monitoring component as an additional virtual function, which runs on a device capable of receiving information from (i) \emph{Function Manager}, on the changes in registration requests, (ii) sensors, either directly or indirectly (cf. \circled{y} in Figure \ref{fig:sch}), and transferring to the \emph{Decision Engine} of the \emph{Master Device}. The context retrieval function 
operates as needed and only reports changes relevant to the \emph{Decision Engine}. 

It is possible for context monitoring to be carried out on different devices, depending on the context induced from the sensor values. Each sensed context is directly or indirectly mapped to a system objective function as shown in Table~\ref{tab:context}.
Context monitoring is an essential element of \sdk. Several implementations of context monitoring exist in the literature \cite{ Beach:2010, seemon}. These can be adapted for \sdk. The context monitor evaluates the cost of obtaining the measures and selecting the appropriate mechanism/sensor with which to obtain this information. It is also possible that the context may be obtained from the device's operating system, if those features at that level are enabled \cite{Vallina-Rodriguez:2011}. 

Since \sdk~is intended to be used across multiple devices and multiple applications, contexts are represented in the common format described as \texttt{(name, value)} pairs.
Context pairs can be defined per-application and per-device in the specific configuration files.  The state of a particular context pair is expressed and stored as an enumerated type or string. The \texttt{value} field could be a) a threshold, b)  ranges (e.g. moderate temperature could be 20-30 C),  or c)  a binary value (e.g. the device is either charging or discharging).  The context \texttt{value} triggers a context change event and orchestrates appropriate flow of events to notify the \emph{Master Device}. 



\subsection{Decision engine}
\label{sec:DecisionEngine}
Initially, at the system bootstrap, the \emph{Decision Engine} module of each device performs the master selection algorithm and elects the \emph{Master Device}. Then the \emph{Decision Engine} module on the \emph{Master Device} executes the function allocation problem (\textsc{fap} (cf. Section~\ref{sec:FAP})), based on received context information from all client devices in the PAN. It determines the mapping of application function registration requests to actual function execution across devices.  

The \textit{Decision Engine} is triggered for a new assignment of functions by 
the \emph{Context Monitoring} module on changes of context. 
First, the decision engine performs the preferred mappings and filters out infeasible function executions by considering the preferences of device, user and application. The preference information is transferred at the context changes. 
Then, each remaining function registration request is mapped to one of the feasible implementations 
as described in Section~\ref{sec:FAP}. 

Then, the \emph{Master Device} creates individual messages for each device which has 
(i) the mapping between function registration requests $r \in R_d$ and the device $d \in D$ selected to execute the function ($R_d \mapsto D$), (ii) the mapping between function executions $v \in V_d$ in the device and each requesting device $d \in D$ ($V_d \mapsto D$). These messages are then transferred to the \emph{Communication Manager} of the \emph{Master Device} in order to communicate to all the devices in the PAN (cf. \circled{g} in Figure \ref{fig:sch}).

\subsubsection{Objective functions}
Although the overall objective of \sdk~is to maximize the user quality of experience, there can be specific objectives for individual users. We have categorized the potential user specific objectives into three groups: i) \emph{Quality}, ii) \emph{Energy}, and iii) \emph{Monetary cost}. An example set of context information that is required by  the \emph{Decision Engine} to achieve a specific objective is summarized in Table~\ref{tab:context}.
We have given the user and/or application developer the ability to configure the optimization preference (via the UI or \emph{Preference APIs}); by default \sdk~is configured to optimize \emph{Energy}. In the case of \emph{Quality}, we assume the user prefers quality of service improvements, e.g., maximize the precision of sensing information or network throughput, over energy consumption of devices and monetary cost. Then, in the case of \emph{Energy}, we presume the user prefers to keep all devices in the PAN active for the longest possible time compromising \emph{Quality} and \emph{Monetary cost}. Finally, if the user opts for \emph{Monetary cost}, the \emph{Decision Engine} makes external communication decisions based on cost per byte information provided by the user (via the UI provided by \sdk). However, achieving one objective does not gurentee the achievement of another objective. 
In order to support different objectives and function categories the \emph{Decision Engine} runs an instance of the function allocation problem (\textsc{fap}) per objective-function category pair (cf. Section~\ref{sec:FAP}).




\subsection{Communication manager}
\label{subsec:cmanager}


This module manages all \sdk~communications. 
There are two types of communication modes in \sdk:  
 i) Inter-device communication and ii) Intra-device communication.

\subsubsection{Inter-device Communication}
\label{subsec:SystemDesign}

Communication among devices in the PAN is performed via Bluetooth or other similar low-powered wireless technologies. We define \sdk~specific message formats rather than using existing data structures (e.g., Java-defined, csv) in order to minimize the amount of data transferred. The defined message formats and descriptions are shown in Appendix A. In addition, the \emph{Communication Manager} aggregates messages and batches data transfers to further limit communication costs.
Inter-device communication in \sdk~is performed in following situations.

 \textbf{(i) During the system bootstrap or after any device joins/leaves the PAN.} All the devices broadcast their own capabilities (e.g., via \texttt{DataAPI} in Android). \emph{Initialization Message} is used in this phase. The main purpose of this message is to announce the device to the PAN along with its supported functions.  

 \textbf{(ii) Changes in the context.} In this phase, a \emph{Context (Sensor) Message} or \emph{Context (Request) Message} is transmitted from client devices to master device when there is a change in any of the context information.

 \textbf{(iii) Once the \emph{Decision Engine} module on the \emph{Master Device} executes the function allocation algorithm.} The \emph{Communication Manager} of the \emph{Master Device} notifies the other devices with the assignments using \emph{Assignment Message}, which contains the $R_d \mapsto D$, and $V_d \mapsto D$ mappings.

 \textbf{(iv) Data transfer from/to other devices in the PAN.} When the requested function is selected to run in a different device than the requesting device, the required data is transferred using a \emph{Data Message}. The \emph{Communication Manager} is designed to aggregate data for different requests to a particular device in order to remove the additional tail energy after each message transmission.

\subsubsection{Intra-device communication}
Intra-device communication involves in data exchanges between the \sdk~and \sdk-enabled applications. \textit{Data Message} format is used for these transfers. There are four methods available for intra-device communication:  \emph{broadcasting, sockets streams, content provider} (specifically in Android), and \emph{service callbacks}. 
In \emph{broadcasting}, all the apps registered with a given broadcast listener receive the broadcasted data, while apps that are not registered with the listener will discard the message. The \emph{socket streams, content provider} and \emph{service callbacks} methods provide one-to-one data transmission instead, where data is directly exchanged between \sdk~and the targeted app. 

Intuitively, the most suitable mechanism for intra-device communication depends on the requested function. For example, in the case of sensing functions where multiple different applications are requesting the same function, the use of \emph{broadcasting} would be efficient.
On the other hand, in case of data transfer towards the Internet where applications have different data to transfer, it should be more efficient if data is transferred directly to the application. We experimentally evaluate the energy efficiency of all these methods in Section~\ref{sec:ExpEval}.


\subsection{Function execution}
\label{subsec:fexec}
This module is responsible for the invocation of functions on the devices selected by the \emph{Decision Engine}. The \emph{Communication Manager} forwards the required function invocations (cf. \circled{h} in Figure \ref{fig:sch}) 
Then, the \emph{Function Execution} module leverages operating system APIs to execute functions. Essentially, \emph{Function Execution} maps the 
function requests made by \emph{\sdk~APIs} to operating system APIs and then invokes the operating system APIs accordingly. The data exchanges in between the \sdk-enabled applications and \sdk~is performed via  
via the \emph{Communication Manager} (cf. \circled{k} in Figure \ref{fig:sch}).

\section{Context-aware Function Allocation}
\label{sec:FAP}

We define  $r_{a,v,d}\in R$ to be the registration for  virtual function $v \in V$, at  device $d \in D$, for application $a \in A$, where $A$ is the set of applications installed in the PAN.
$R_v\subset R$ and $D_v \subset D$ represents the set of registrations and the list of devices providing implementations respectively for a given virtual function $v$. Thus, the objective of the \emph{context-aware function allocation} is to map each function registration request to its chosen implementation, i.e.  $R_v \mapsto D_v$, to optimize the total cost of executing the all requested functions.

\subsection{Function costs}
The function costs are related to the usability objective of the system, i.e. monetary, quality and/or energy, which can be defined either by the app developer or the user. For each objective, there are two types of costs associated with each function request and its implementations; 1) communication costs and 2) implementation costs. If the objective is to optimize the monetary costs,
internal communication (e.g., Bluetooth), can be considered as zero. On the other hand, if the objective is to optimize the energy consumption of the devices, communication is not negligible. Usually, 
local mapping incurs zero cost.
For the same function, the implementation cost can be different for multiple devices, for example, the energy cost of activating the WiFi network interface compared to the total battery capacity on the smartphone is lower than on the smartwatch. 

The implementation function cost  $v$ in device $d$ is represented as $f_{v,d} \in F_v$. Similarly, the communication costs between a given function registration request $r_{a,v,d}$ and an implementation on device $d$ are denoted as $c_{r,d}\in C_r$.

\subsection{Problem formulation} 
We first define a binary variable $m_{r,d}$ where $m_{r,d}=1$, if function registration request $r_{a,v,d} \in R_v$ can be mapped to implementation on device $d \in D_v$, and $m_{r,d}=0$ otherwise, depending on the current context of the user and the device. For instance, even if the GPS sensor is implemented on the device $d_1 \in D_v$, it may not be able to map $d_1$ with any request if the current remaining battery capacity on $d_1$ is below the threshold. If no function implementation is available for a particular function registration request, we remove that function from the problem formulation. That makes for all considered functions $\sum_{\forall d \in D_v} m_{r,d}=1; \forall r \in R_v$. Given the set function registrations $R_v$ and function implementations $D_v$ and the associated costs $f_{v,d}$ as input, the optimal \textsc{function allocation problem} (\textsc{fap}) can be formulated as follows:
\begin{equation}\label{eq:test}
\textrm{Minimize}  \left(\sum_{d \in D_v}y_{d} \cdot f_{v,d} + \sum_{r \in R_v} \sum_{d \in D_v} x_{r,d} \cdot c_{r,d}\right)
\end{equation}
Subject to:
\begin{align*}
1. \quad &\sum_{d \in D_v} x_{r,d} =1; &\forall r \in R_v\\
2. \quad & m_{r,d}  \ge x_{r,d}; & \forall r \in R_v, \forall d \in D_v\\
3. \quad & y_{d}  \ge x_{r,d}; & \forall r \in R_v, \forall d \in D_v\\
4. \quad & y_{d}, x_{r,d} \in \{0,1\}; &\forall r \in R_v, \forall d \in D_v
\end{align*}
The sets of $x_{r,d} \in X$ and $y_d \in Y$ would be the solution of the \textsc{fap}. $x_{r,d}=1$ if the function registration request $r \in R_v$ is assigned to the device $d \in D_v$ and $y_d=1$ if the device $d$ is required to be activated to satisfy certain requests. Only mappable implementations will be assigned and each function registration request will be mapped to an implementation.

\subsection{Solution to function allocation}
When $m_{r,d}$ is given $\forall r \in R_v, \forall d \in D_v$, it is trivial to show that \textsc{fap} is equivalent to \textsc{Uncapacitated Facility Location (ufl)} problem where every function implementation $D_v$ is a facility with $f_{v,d}$ facility opening cost and every function registration request $R_v$ corresponds to a customer associated with $c_{r,d}$ service cost. It immediately follows that \textsc{fap} is also an \emph{NP-Hard} problem. However, there are number of approximation algorithms for the well-studied \textsc{ufl} problem. We build on the approximation algorithm proposed by Williamson and Shmoys~\cite{approxbook} to take into account the use of valid mappings ($m_{r,d}$) after context aware constraints. The iterative greedy solution to \textsc{fap} is described in Algorithm~\ref{algo:fap}.

\begin{myalgorithm}[8.2cm]{\sc fap$(R_v, D_v, m, f, c)$}
\begin{footnotesize}
\label{algo:fap}
 \STATE  $S$ $\leftarrow$ $R_v$
 \STATE  $X$ $\leftarrow$ $\emptyset$ 
 \WHILE {$S \neq \emptyset$}
     \STATE Select $v \in D_v$ and $P \subseteq S$ s.t. $\forall p \in P: m_{r,d}=1$ \\ that minimize $\frac{f_{v,d} + \sum_{p \in P } c_{r,d}}{|P|}$
     \STATE $S$ $\leftarrow$ $S -P$; $f_{v,d}=0$
     \STATE $(R_v \mapsto D_v$) $\leftarrow$ $(R_v \mapsto D_v$) $+$ ($P$ $\mapsto$ $v$)
  \ENDWHILE
  \RETURN assignment $\sigma: R_v \mapsto D_v$
\end{footnotesize}
\end{myalgorithm}

The algorithm  iteratively selects a function implementation and the valid registrations associated to it.
Assigned registrations are then removed from the problem and the implementation cost set to 0. At each iteration, an implementation is selected as to minimize the total cost of function registrations that will be associated to the implementation. The algorithm can be efficiently realized by maintaining  the list of registrations not yet satisfied for each implementation in increasing cost order. 
\textsc{fap} is performed for each type of function separately, and aggregate the solutions to get the final solution. This further increases the efficiency of \textsc{fap}.


\section{Realization of AFV Framework}
\subsection{Implementation}
\label{sec:implementation}


Our prototype realizes \sdk~as a library that can be compiled into an application and as a stand-alone user-level application.
For simplicity and without loss of generality, it is assumed that all Tier 1 devices in the PAN run the same OS (e.g., Android). 
The library, \textsf{\sdk~lib}, provides access to the \emph{\sdk~APIs} for the developer once imported to application. To support \sdk~services to multiple applications in parallel at the user-level, the other components of \sdk~are implemented in a standalone application, \textsf{\sdk~app}. At the time of installation of an \sdk~enabled application, it checks whether the \textsf{\sdk~app} is already installed. 
If not, it initiates the installation of the \textsf{\sdk~app}. 

The communication between \sdk-enabled devices is implemented using \texttt{MessageAPI} and \texttt{DataAPI} as  Android \texttt{Services}. Context monitoring is also implemented as an Android \texttt{Service}. A \emph{Master Device} is selected among the devices in the network and it runs \textsc{fap} algorithm described in Section \ref{sec:FAP} to check for optimal function placement.

Device arrival/departure is monitored using the \texttt{onCapabilityChanged} method. 
When an \sdk-enabled device joins/leaves the PAN, all \sdk-enabled devices send \emph{Initialization Messages} via \texttt{DataAPI} 
to all the connected devices.
In addition, \emph{Context Monitoring} runs in the background and collects information such as battery level, charging status, moving status, connected network, and link speed. When a context of a device changes, it reports the change to the \emph{Master Device} by sending a \emph{Context (Sensor) Message} via \texttt{MessageAPI}. This context change will trigger the \emph{Decision Engine} of the \emph{Master Device} to make new decisions. 

\begin{figure}[htb]
	\centering 
	\includegraphics[width=0.5\textwidth]{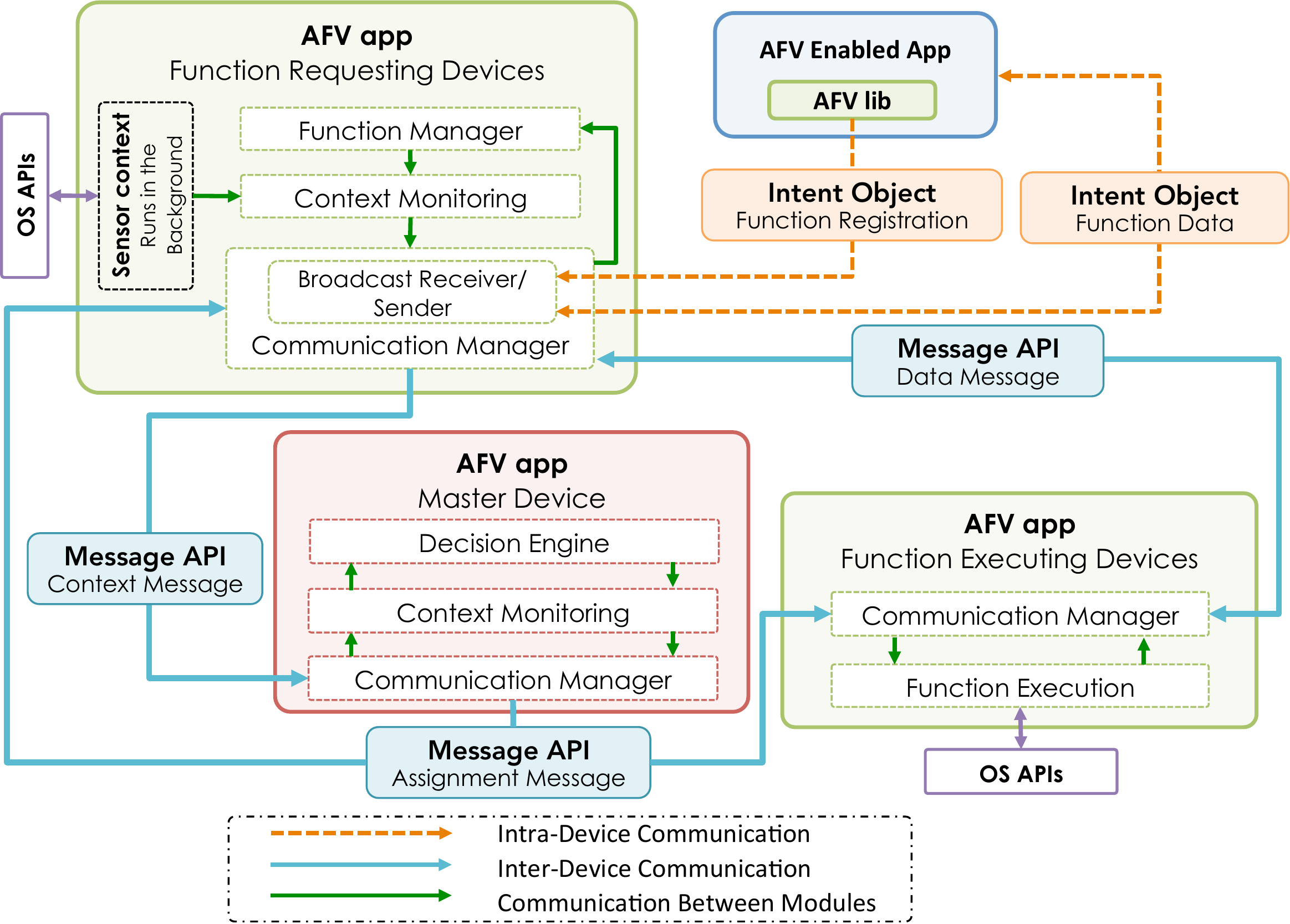}
	\vspace {-3mm}
	\caption{Message flow in the implementation.}
	\vspace {-3mm}
	
	\label{fig:implementation}
\end{figure}

Also, as shown in Figure \ref{fig:implementation}, once an application registers a function (e.g., \texttt{\sdk registerListener(this, Sensor.TYPE \_ACCELEROMETER), AFVHTTPGet}), \textsf{\sdk~lib} sends a broadcast with an \texttt{Intent} about the function registration to \textsf{\sdk~app}. The \texttt{BroadcastReceiver} in the \textsf{\sdk~app} transfers the request to the \emph{Function Manager}. Then, it is identified as a context change and reports to the \emph{Master Device} by sending a \emph{Context (Request) Message} via \texttt{MessageAPI}. Then \emph{Master Device}'s \emph{Decision Engine} checks for the optimal placement of the function. 

The \textit{Decision Engine} implements the \textsc{fap} algorithm by using efficient ordered data structures (i.e., \texttt{TreeMultimaps} and \texttt{ArrayLists}). \textit{Function Execution} registers the function using Android APIs and returns data in a \emph{self-defined data format} to the \textit{Communication Manager} (\emph{Data Message}). The \textit{Communication Manager} aggregates all the \emph{Data Messages} per device and sends it via \texttt{MessageAPI}. Once the \emph{Data Message} is received by a device, the \textit{Communication Manager} broadcasts the data stream. Each app that has a registered \texttt{listener} for the function will receive the data stream. 

In Section \ref{sec:ExpEval}, we use \textsf{\sdk~lib} and \textsf{\sdk~app} to evaluate the practical feasibility of \sdk~and user benefits with use cases.


\subsection{Experimental Calibrations}
\subsubsection{Energy costs of functions}
For the \textsc{fap} algorithm to allocate functions optimally, the \emph{Function Manager} should contain a list of supported functions and their associated costs. We measured energy consumption of several application functions for sensing, communication,\footnote{Only WiFi \textit{receive} measured, \textit{transmit} is between 20\% and 30\% higher \cite{modeling-wifi}.} and processing. We use a  smartphone running Android 6.0 and LG Watch Urbane running Android 5.1.1 with 2300 mAh and 410 mAh battery capacities respectively for all experiments. The functions are reported in Table~\ref{tab:functionscosts}.

\begin{table}[ht]
\caption{Energy cost associated to each function }

\centering
\scriptsize
\begin{tabular}{|l|c|c|}
\hline
\multirow{2}{*}{\textbf{Function}} & \multicolumn{2}{|c|}{\textbf{Energy cost}} \\
\cline{2-3}
& \textbf{Smartphone} & \textbf{Smartwatch} \\
\hline\hline

 \multicolumn{3}{|l|}{\textbf{Sensing (Speed (NORMAL - UI - GAME - FASTEST ) [mJ/s])}}\\\hline
Accelerometer & 5.01 - 13.28 - 34.46 - 77.71 & 9.52 - 24.74 - 57.61 - 168.40 \\\hline
Gyroscope& 11.71 - 20.33 - 36.44 - 80.15 & 16.23 - 33.34 - 60.44 - 181.90 \\\hline
Magnetometer& 8.12 - 15.45 - 28.46 - 28.28 & 17.04 - 30.21 - 57.82 - 79.73 \\\hline\hline
\multicolumn{3}{|l|}{\textbf{Connectivity (Per Byte [mJ/B] - High Power Idle [mJ] - Low Power Idle [mJ])}}\\\hline
Bluetooth& 0.0095 - 305 - 300 & 0.0024 - 126.07 - 64.23 \\\hline
WiFi& 0.00054 - 66 - N/A & 0.00039 - 50 - N/A\\\hline\hline
 \multicolumn{3}{|l|}{\textbf{Processing (Per Byte [mJ/B])}}\\\hline
Compression&0.01&0.0004\\\hline
Encoding&0.00026&0.00025\\\hline
\end{tabular}
\label{tab:functionscosts}
\end{table}

The energy consumption for each function is obtained with a Monsoon power monitor\footnote{\url{https://www.msoon.com/LabEquipment/PowerMonitor/} } directly connected to each device via USB. Energy usage is obtained by integrating the instantaneous power values
calculated using current and voltage measurements from the USB interface sampled at 0.2~ms time intervals. The experiment energy usage is computed by deducting the fixed energy of the background processes from the total energy consumed. Sensing energy is measured for multiple sampling frequencies that are offered by Android by default.\footnote{\url{http://developer.android.com/reference/android/hardware/SensorManager.html}} These energy cost values are used in the experimental validation of \sdk~architecture and to show user benefits.

\subsubsection{Intra-device communication modes}

\begin{figure}[tb]
	\centering  
	\subfigure[With one external app.]{\label{fig:interAppComm}\includegraphics[width=0.3\textwidth]{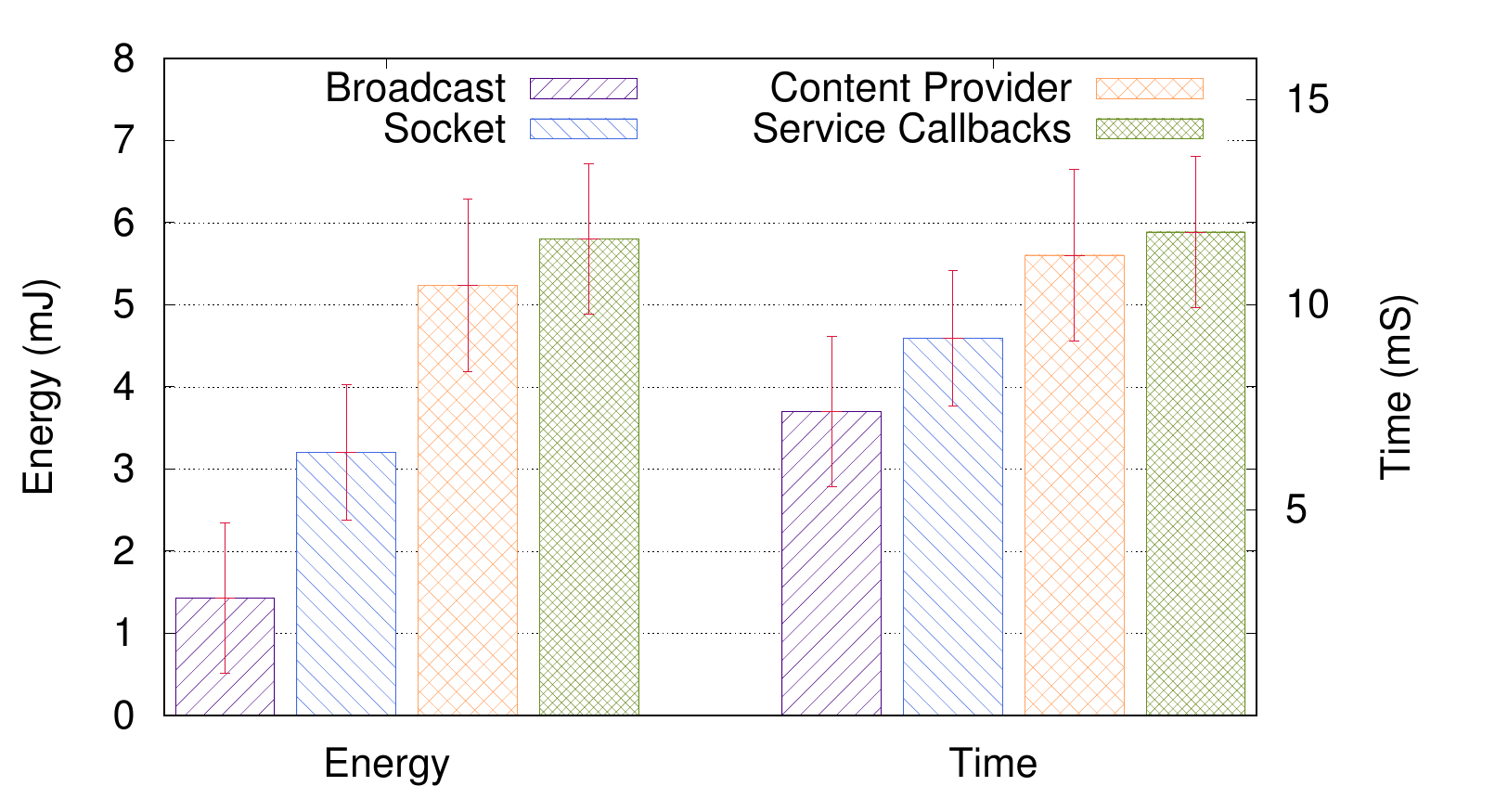}}
	\subfigure[Broadcasting to  multiple apps requesting same data.]{\label{fig:Broadcast}\includegraphics[width=0.15\textwidth]{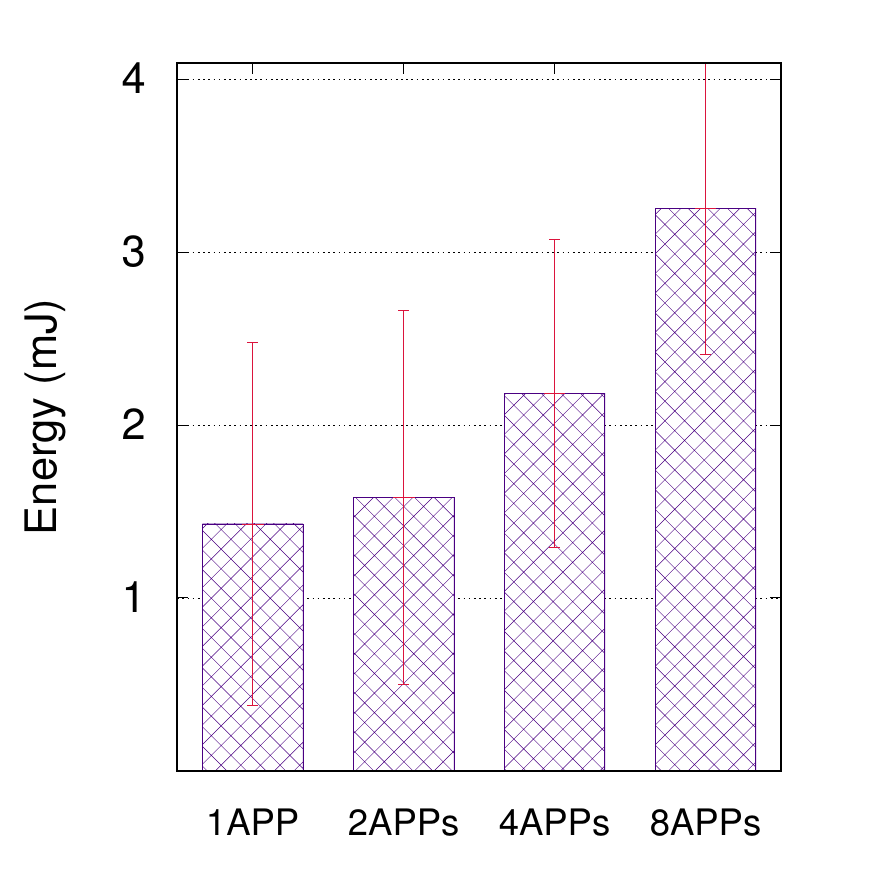}}
	\vspace {-2mm}
	\caption{Impact of intra-device communication modes.}
	\vspace {-2mm}
	
	\label{fig:Adaptation} 
\end{figure}

As mentioned in Section~\ref{subsec:cmanager}, there are four potential modes for intra-device communication. 
We experimentally evaluate the energy efficiency of four methods (\emph{broadcasting, socket streams, content provider, service call-backs}) in the case of \sdk~message exchanges. Figure~\ref{fig:interAppComm} shows the results for different intra-device communication methods in transferring 2.5~KB of data from \sdk~to an app. The measurements are done in an Android smartphone. As expected, intra-device communication has much lower energy usage than inter-device communication (cf. Table~\ref{tab:functionscosts}). In this particular case, smartphone inter-device communication consumes $\sim$23~mJ without tail energy ($\sim$650~mJ with tail energy) and intra-device communication consumes $\sim$1~mJ-6~mJ.

Moreover, we observe that \emph{broadcasting} requires the minimum energy for transactions even in case of a single application. Therefore, we further experiment with the case where up to eight external apps requesting the same data from \sdk~architecture. Figure~\ref{fig:Broadcast} illustrates that energy consumption of \emph{broadcasting} to eight apps is still lower than any other method of intra-app communication for a single one app as shown in Figure~\ref{fig:interAppComm}.  
We thus select \emph{broadcasting} as the intra-device communication mode in \sdk.

\section{Performance Evalaution}
\label{sec:ExpEval}


We first evaluate the performance of the \textsc{fap} algorithm with data driven simulations. 

\subsection{Evaluation of the FAP Algorithm}
\label{evaluation}

\begin{figure}[t]
	\centering  
	\subfigure[The impact of cost ratio $F_v/C_r$.]{\label{fig:optimcost}\includegraphics[width=0.45\textwidth]{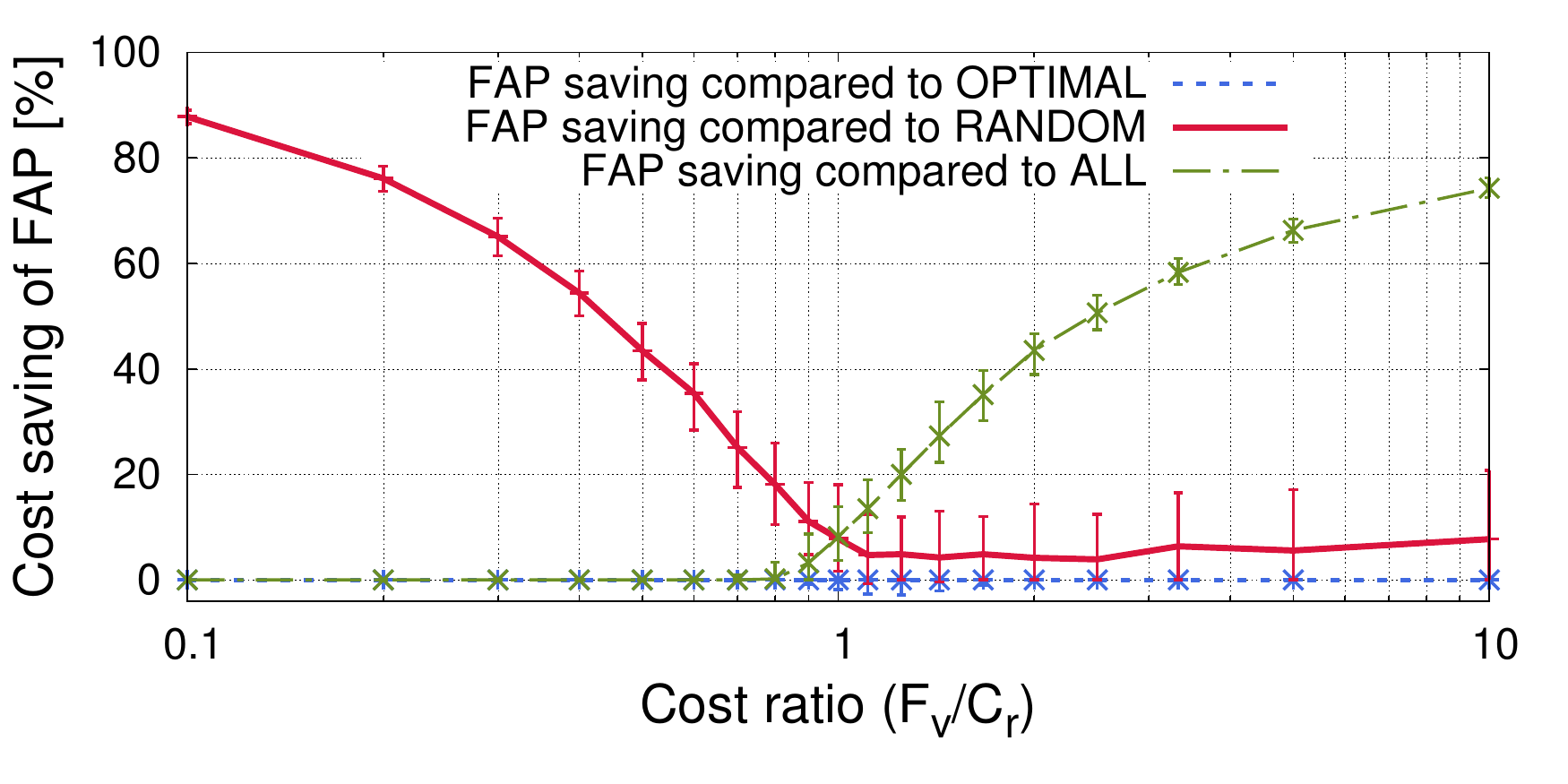}}
	\subfigure[The impact of no. of functions, $F_v/C_r=1$.]{\label{fig:optimscaling}\includegraphics[width=0.45\textwidth]{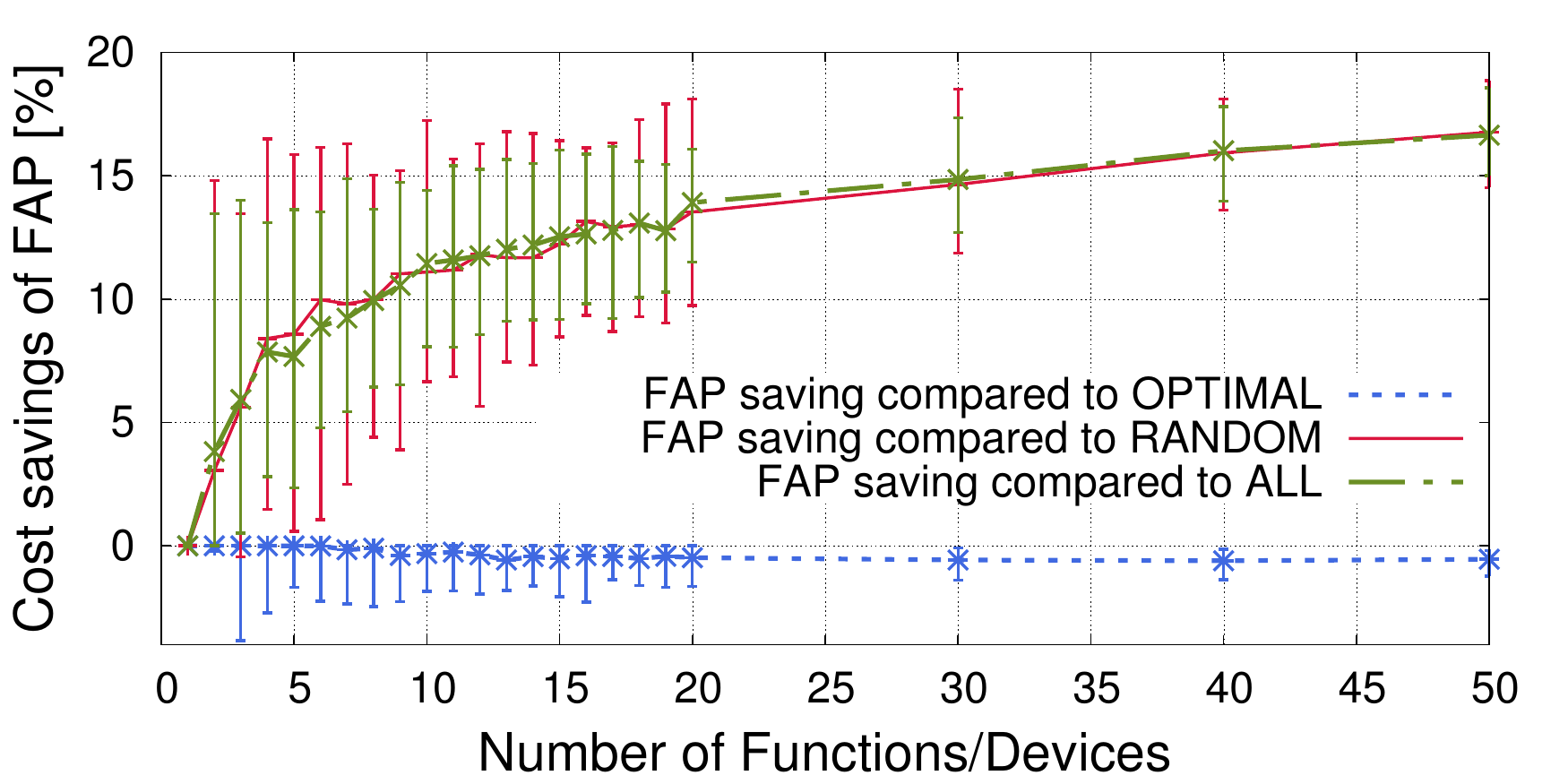}}
	\caption{Efficiency, accuracy and robustness of \textsc{fap} algorithm.}
	\label{fig:simulations} 
	\vspace {-5mm}
\end{figure}

\subsubsection{Evaluation methodology}
We developed a custom simulator to analyze the effectiveness of the \emph{Decision Engine}, inparticular, \textsc{fap} algorithm. We compared \sdk~function allocation against three strategies: \vspace{-2mm}
\begin{itemize}
\item \textsc{manual}: User assignment of functions in a static manner, e.g., \textsf{MyFitnessCompanion} \cite{Kolamunna2016}.
\item \textsc{all}: Running functions on all available devices in parallel, which is one of the common strategies in today's wearable applications, e.g., \textsf{UP} \cite{Kolamunna2016}.
\item \textsc{optimal}: Function allocation, using the optimization problem solver Gurobi.\footnote{\url{https://www.gurobi.com}}
\end{itemize}
\vspace{-2mm}
We assume that costs of executing a function on devices is normally distributed, with a standard deviation $\sigma=0.1 \times \mu$ where $\mu$ is the average value. We change $\mu$ to obtain multiple cost values to evaluate the performance of \textsc{fap}.

\subsubsection{Efficiency and robustness of the FAP algorithm }
Figure~\ref{fig:optimcost} shows the cost reduction obtained when using the \textsc{fap} algorithm with respect to \textsc{manual}, \textsc{all} and \textsc{optimal} as a function of the ratio of function implementation cost ($F_v$) to communication cost ($C_r$). Communication costs are incurred for any transmission of an individual sensor stream to the device executing the application.

We consider 5 active wearable devices in a PAN. Intuitively, if the communication cost is too high, it is more efficient to execute the function on each device, resulting in parallel apps with no coordination. This is reflected in the region where $F_v/C_r < 1$. Under these conditions \textsc{all} performs as well as \textsc{optimal} and \textsc{fap}. However, \textsc{fap} significantly reduces the cost compared to \textsc{manual} selection. As $F_v/C_r$ increases the significance of the communication cost decreases. Thus, executing a function in all devices becomes inefficient as there is potentially a device with a very low relative function execution cost. Since there is a 1/5 chance of selecting the right device, \textsc{manual} performs comparatively well with a high standard deviation. \textsc{fap} performs equally well (error is less than 1\%) compared to \textsc{optimal} irrespective of the $F_v/C_r$ value.

Figure~\ref{fig:optimscaling} shows that \textsc{fap} increases its cost savings compared to both \textsc{manual} and \textsc{all} along with the number of functions when $F_v/C_r=1$. Furthermore, \textsc{fap} accuracy does not vary significantly compared to \textsc{optimal} (error is about 2-3\%). Overall, Figure \ref{fig:simulations} shows that the  \textsc{fap} algorithm is often able to map the function registration requests to the optimal device for executing the function providing significant cost savings.

\subsection{Evaluation of prolong system uptime}
We analyze \sdk~effectiveness by considering system uptime (i.e. time until at least one device drains out its battery) as an example of the quality metric. 
First, we evaluate the system uptime varying the power status of devices with simulations. Then, we augment simulation results by conducting experiments with real-devices. We consider a smartphone and a smartwatch for both simulations and experiments.

\subsubsection{Evaluation with simulations}

\begin{figure}[t]
\centering 
\subfigure[Battery drain profiles]{\label{fig:lifetimesimple}\includegraphics[width=0.45\textwidth]{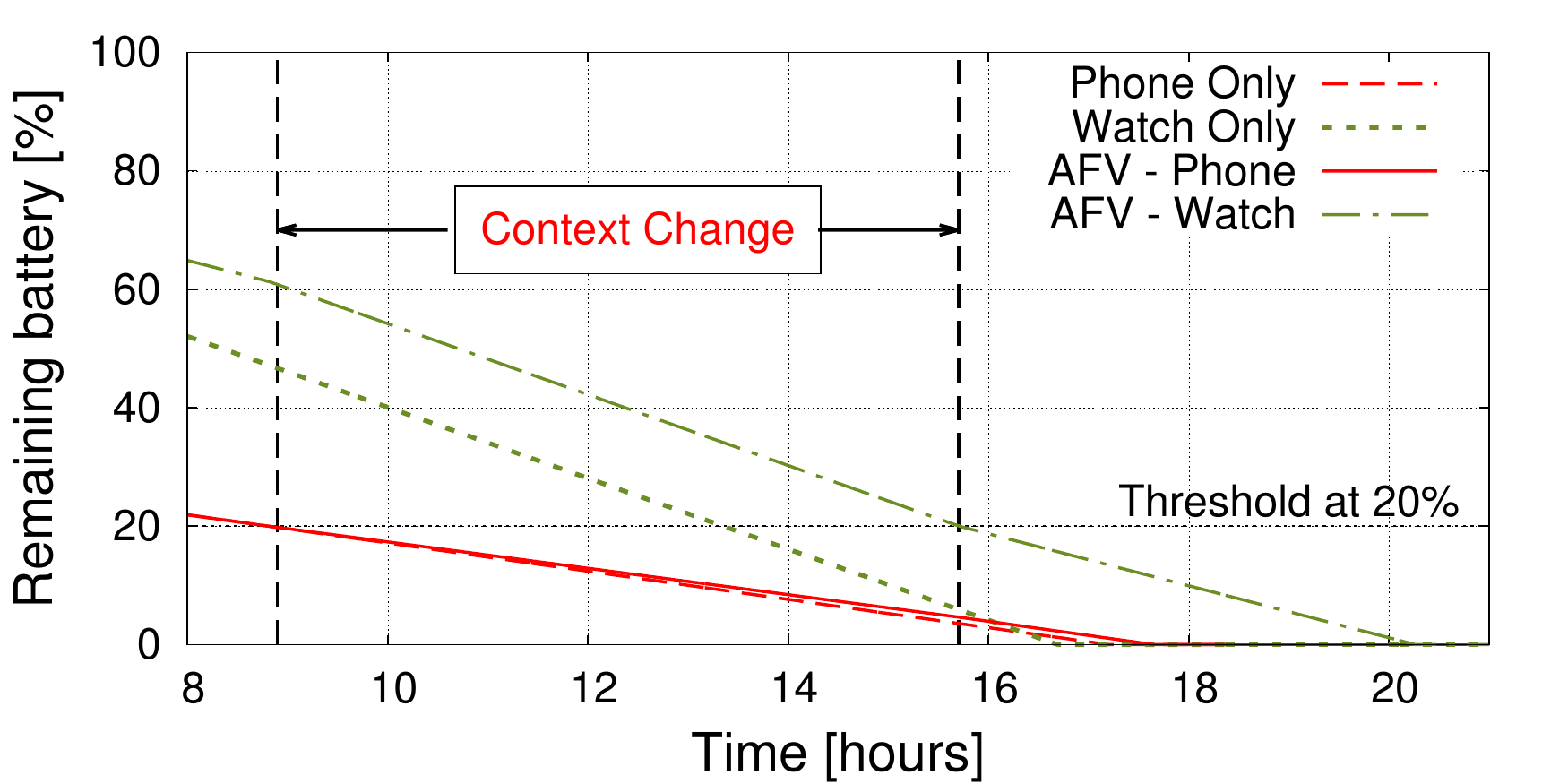}}
\subfigure[Percentage increase in system uptime.]{\label{fig:lifetimeenergy}\includegraphics[width=0.45\textwidth]{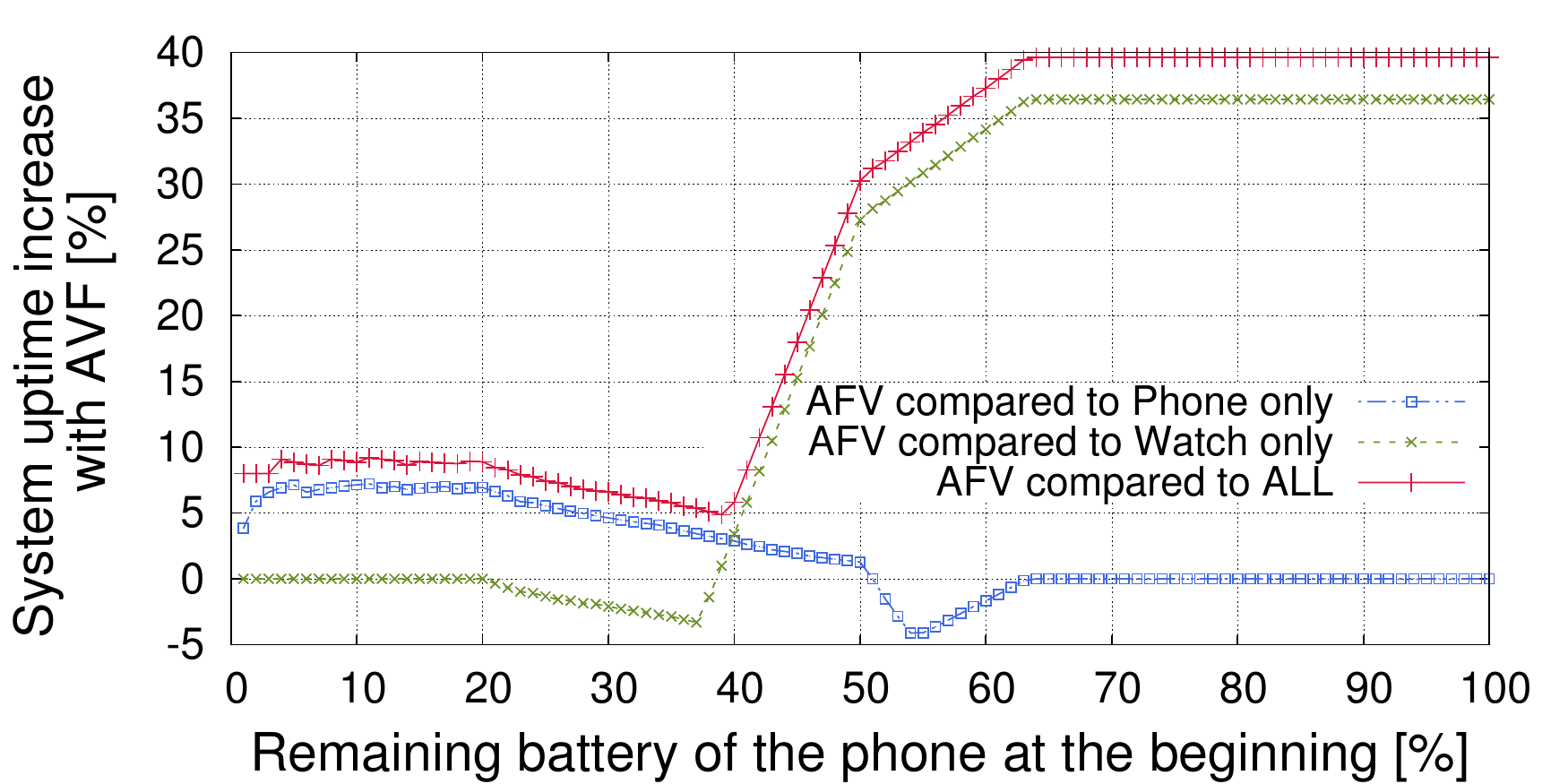}}
\caption{The impact of \sdk~on system uptime.}
\label{fig:lifetime} 
\vspace {-7mm}
\end{figure}

Uptime of a device depends on its remaining battery percentage (i.e. State of Charge (SoC)) and current energy usage. 
To simulate typical user behaviour, we assume the smartphone battery would completely drain in two days linearly and the smartwatch would last only one day. We consider the ``sensing accelerometer in FASTEST speed" function and 60 second data synchronization frequency: the \emph{Decision engine} takes decisions to maximize 
system uptime.
We use measurements in Table~\ref{tab:functionscosts} to derive energy consumption for the functions. As an example, for sense only on smartphone (Accelerometer FASTEST speed) and data synchronization frequency of one minute (70KB of data), we can get the energy consumption per minute from Table~\ref{tab:functionscosts} as {$f+c$=(77.71*60)+((0.0095*70000)+305+300)=5932mJ}.

Figure \ref{fig:lifetimesimple} illustrates the battery drain profile for \textsc{manual} selection, i.e. sense only on the smartphone or on the smartwatch, and when \sdk~is running. Due to a lower relative impact on the smartphone, \sdk~selects the smartphone as the sensing device if the smartphone has sufficient SoC. However, if the smartphone's SoC drops below 20\% (context change), the \textit{Context Monitor} triggers the \emph{Decision Engine} and sensing switches to the smartwatch if the smartwatch has sufficient SoC (Figure \ref{fig:lifetimesimple}). To show this context change, we consider the following initial conditions: smartphone - 45\% SoC, smartwatch  100\% SoC. The smartwatch uptime increases by approximately 2 hours compared to sensing on the smartwatch. The gain for the smartphone is approximately 1/2 hour compared to only sensing on the smartphone.

Since uptime gain is dependent on the initial SoC of devices, in Figure~\ref{fig:lifetimeenergy} we change the initial smartphone SoC.  If the smartphone remaining SoC is greater than 60\% at the beginning, \sdk~increases the system uptime between 35-40\% compared to sensing on the smartwatch and on both devices. Due to sufficient battery capacity on the smartphone, \sdk~selects the smartphone most of the time. As a result, \sdk~does not increase the uptime compared to sensing only on the smartphone. \sdk~may marginally reduce the system uptime when the SoC of one or more devices drops below the threshold. To minimize the application energy consumption, while respecting user preferences, the \emph{Decision Engine} selects the only available device or the most energy efficient when both are under the threshold, although this solution may reduce system uptime. This can be observed when initial Phone SoC is approximately 35\% and 55\%.

\subsubsection{Evaluation with experiments}

\begin{figure}[t!]
	\centering 
	\subfigure [Energy consumption of different message passing phases.]{\label{fig:sysEvaluation} \includegraphics[width=0.45\textwidth]{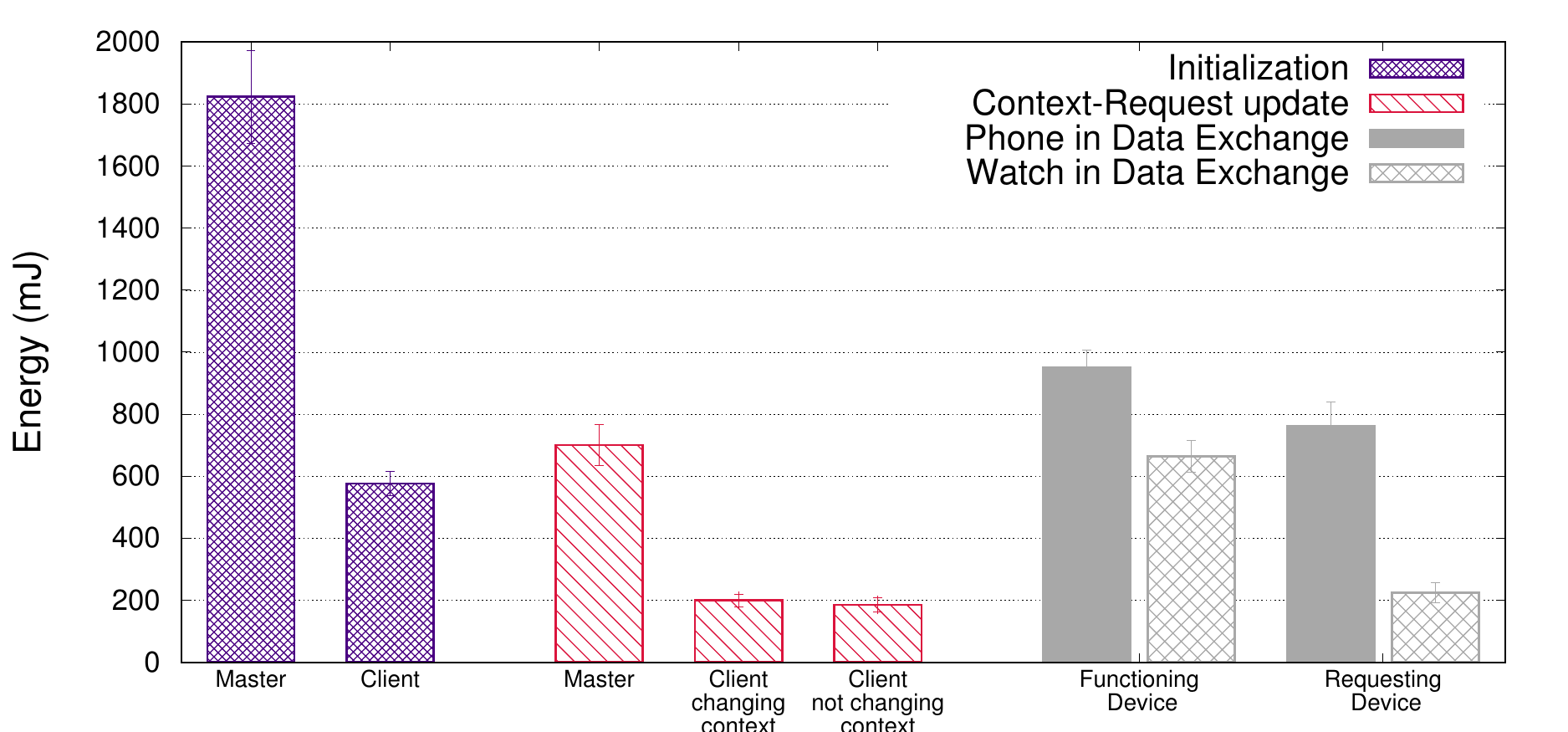}}

	\subfigure[Energy consumption with and without AFV when each device is requesting for different functionalities.]{\label{fig:adapt_SOC1}\includegraphics[width=0.44\textwidth]{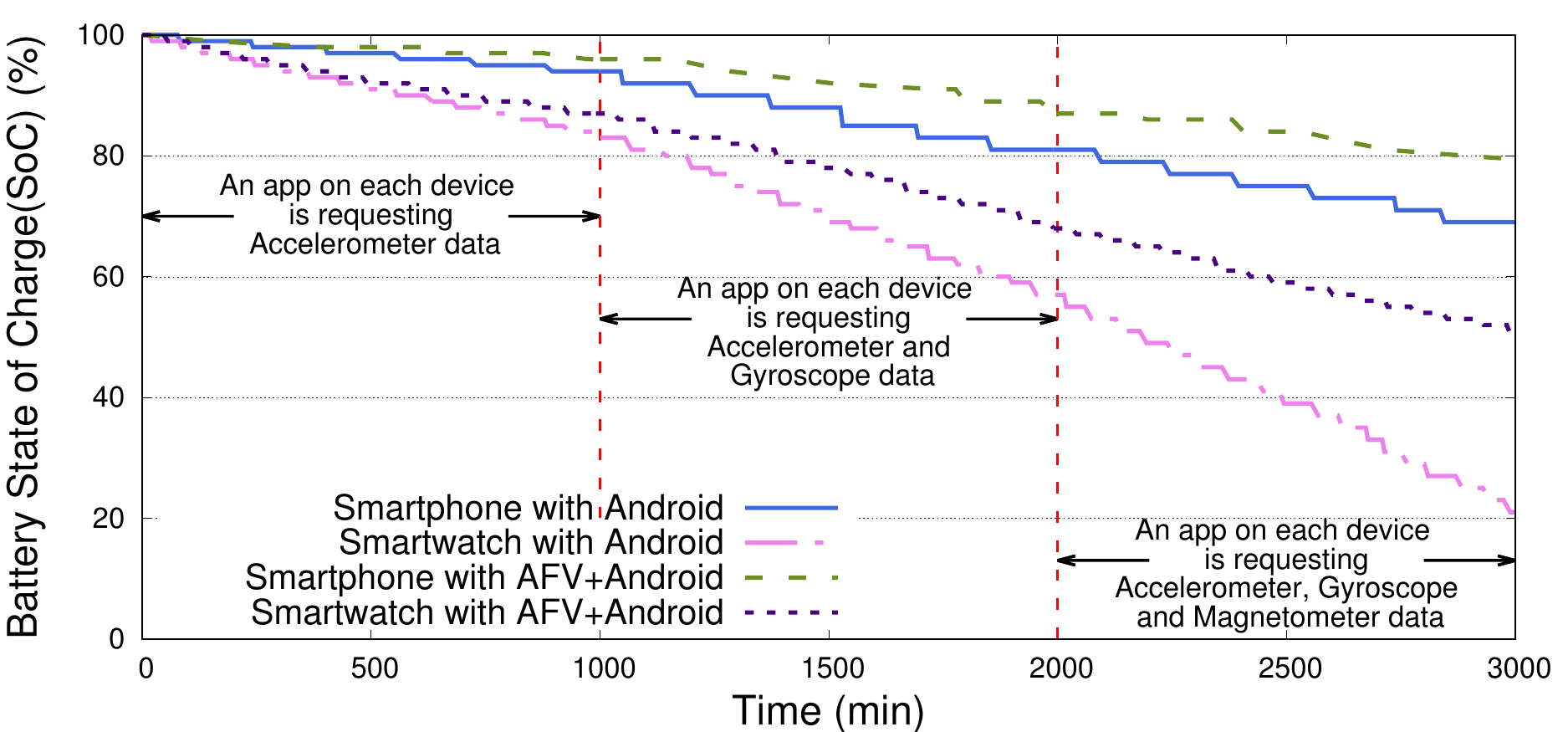}}
	\caption{Energy consumption of the system.}\vspace{-4mm}
	\label{fig:exp} 

\end{figure}

Next, we quantify the energy consumption of the devices with and without \sdk~experimentally. 
We installed \sdk~on an Android smartphone and a smartwatch. 
Without \sdk, we use counterpart applications that are installed on both devices. 
We consider the ``sensing accelerometer in NORMAL speed" function and 60 second data synchronization frequency.
Most current apps select both devices to perform a certain functionality and then exchange data \cite{Kolamunna2016}. Therefore, we selected \textsc{all} function allocation strategy. 

Figure \ref{fig:sysEvaluation} shows the measured energy consumption for each type of message passing in \sdk~(cf. Section~\ref{subsec:cmanager}). The energy requirement for group formation (\emph{Initialization}) is $(0.6*(n-1) +1.8)$J that is much lower compared to the group formation energy in \cite{Hemminki2013}. 
Figure \ref{fig:adapt_SOC1} shows the reduction of battery SoC in smartphone and smartwatch. 
Using \sdk~achieves lower energy usage by approximately 3 times for one function request, 
despite the additional energy consumption of \sdk~(e.g., \emph{Initialization}, \emph{Context Monitoring}). 
Moreover, the battery SoC decreases much faster as the number of functions increases, especially without \sdk.
Thus more energy is saved with \sdk when the number of function requests increases.



\subsection{Experimental validation of \sdk-enabled PAN}
\label{sec:usecases}

\begin{figure}[tb]
	\centering 
	\includegraphics[width=0.45\textwidth]{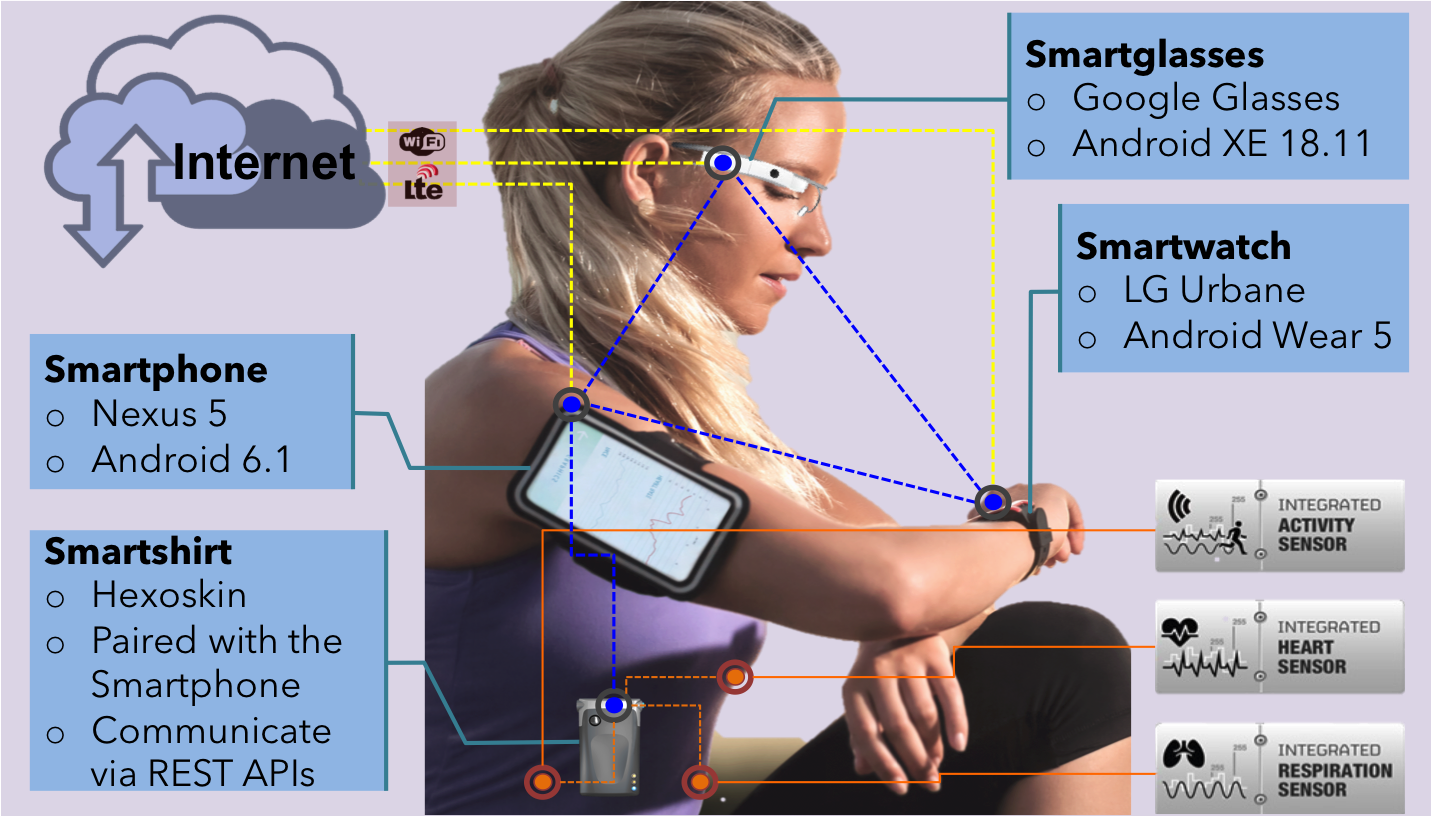}
	\vspace {-3mm}
	\caption{Experimental Setup.}
	\vspace {-3mm}
	
	\label{fig:setup}
\end{figure}

We now present experimental results to show quantitative benefits of \sdk. As shown in Figure \ref{fig:setup}, we consider a PAN consisting of 4 devices (3 Tier 1 devices and a Tier 2 device). The Tier 1 devices (smartphone, smartwatch and smartglasses\footnote{https://developers.google.com/glass/}) are connected with each other via Bluetooth and each have Internet connectivity. The Tier 2 device (smartshirt\footnote{https://www.hexoskin.com/}) consists with three different sensor types and paired with the smartphone. We developed an \sdk~enabled fitness tracking application requiring accelerometer and heart rate information to be uploaded to Internet servers, that is similar to the applications previously identified \cite{Kolamunna2016},
 that is the current popular health and fitness applications. The app was installed in the smartphone, smartwatch and smartglasses.


We investigated five scenarios, to investigate the benefit of using \sdk. These experiments evaluated three main objectives described in Table~\ref{tab:context}. 
For the first objective of achieving the maximum functional quality, we conducted two experiments, one to achieve the best information quality, and the other to achieve the maximum network throughput. 
For the objective of energy utilization, we examined how the usage of \sdk~extends the device uptime. Finally, we examined the case of minimizing the monetary cost of data usage.


\subsubsection{Maximizing the Functional Quality}

\textbf{(1) Maximizing the precision of fitness/health tracking.}  

\begin{figure*}[tb]
	\centering 
	\subfigure[Accelerometer data received by an \sdk-enabled application installed in smartphone.]{\label{fig:new_usecase1}\includegraphics[width=0.32\textwidth]{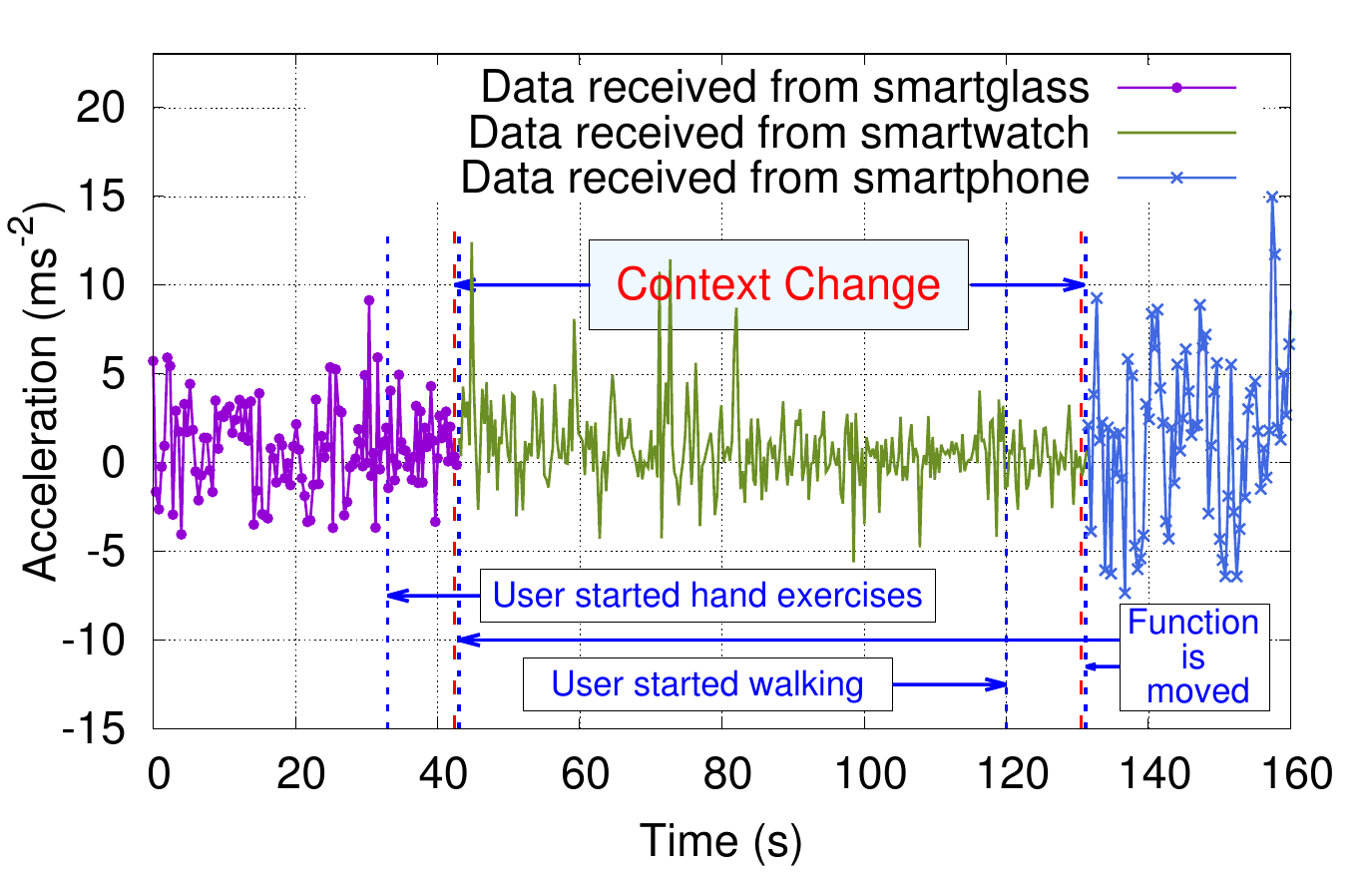}}
	\subfigure[Heart rate data received by an \sdk-enabled application installed in smartwatch.]{\label{fig:new_usecase2}\includegraphics[width=0.32\textwidth]{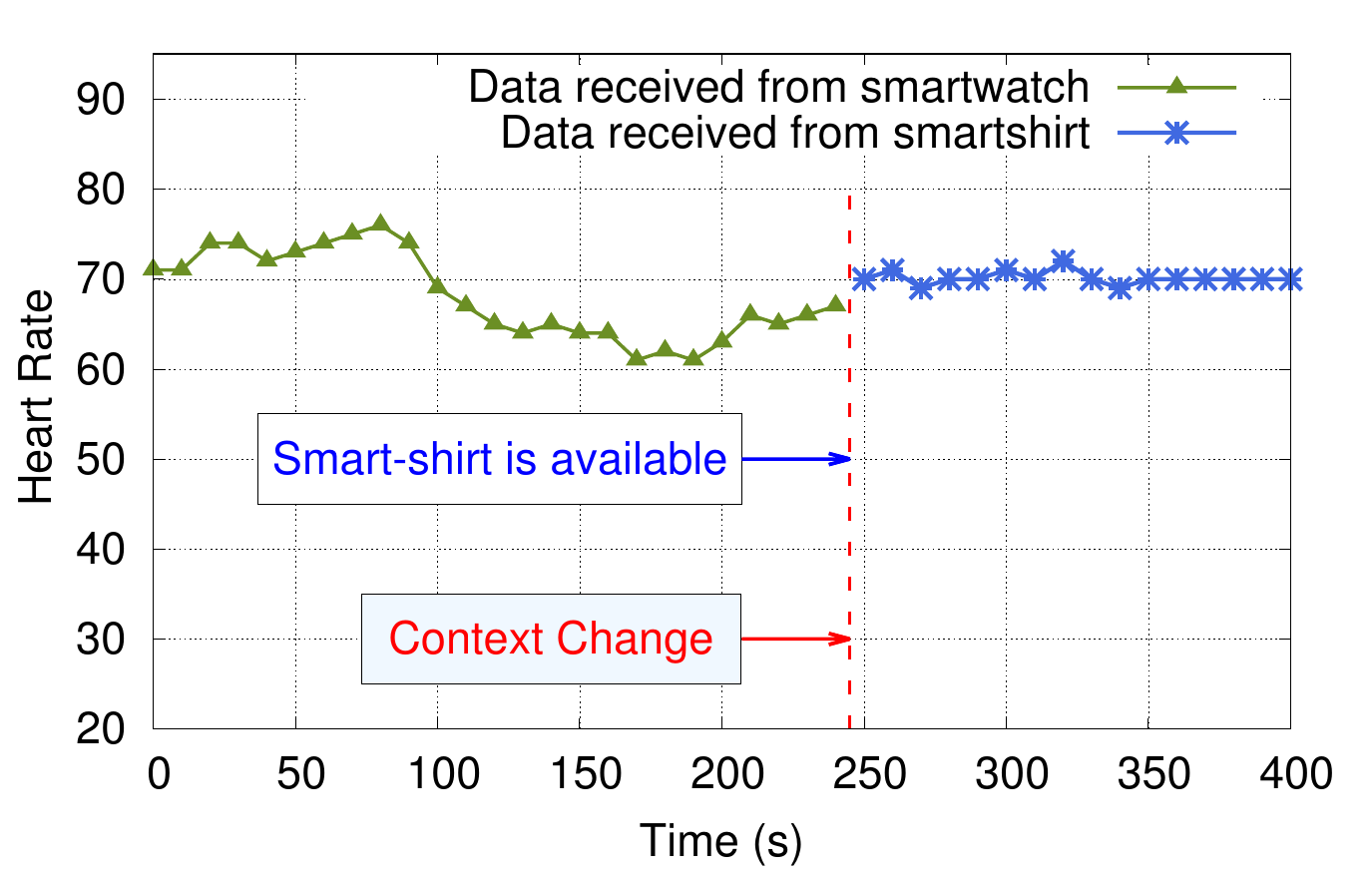}}
	\subfigure[Dynamically switching to the higher data rate network for data uploading.]{\label{fig:adapt_BW}\includegraphics[width=0.32\textwidth]{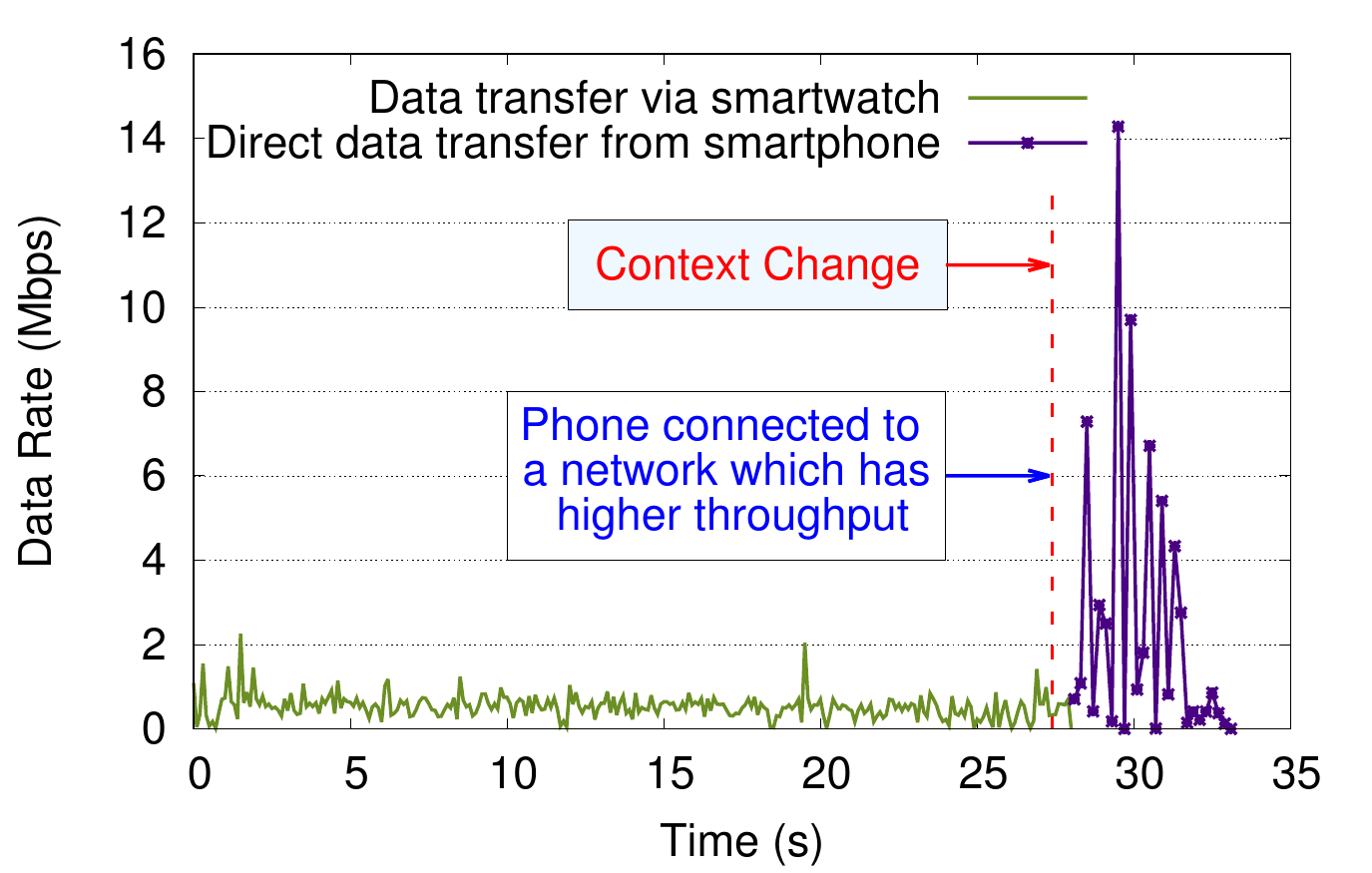}}
	\caption{Maximizing the functional quality with \sdk.}
	\label{fig:lifetime} 
	\vspace {-5mm}
\end{figure*}

\begin{figure*}[tb]
	\centering 
	\subfigure[Energy consumption profiles for devices in the PAN.]{\label{fig:adapt}\includegraphics[width=0.64\textwidth]{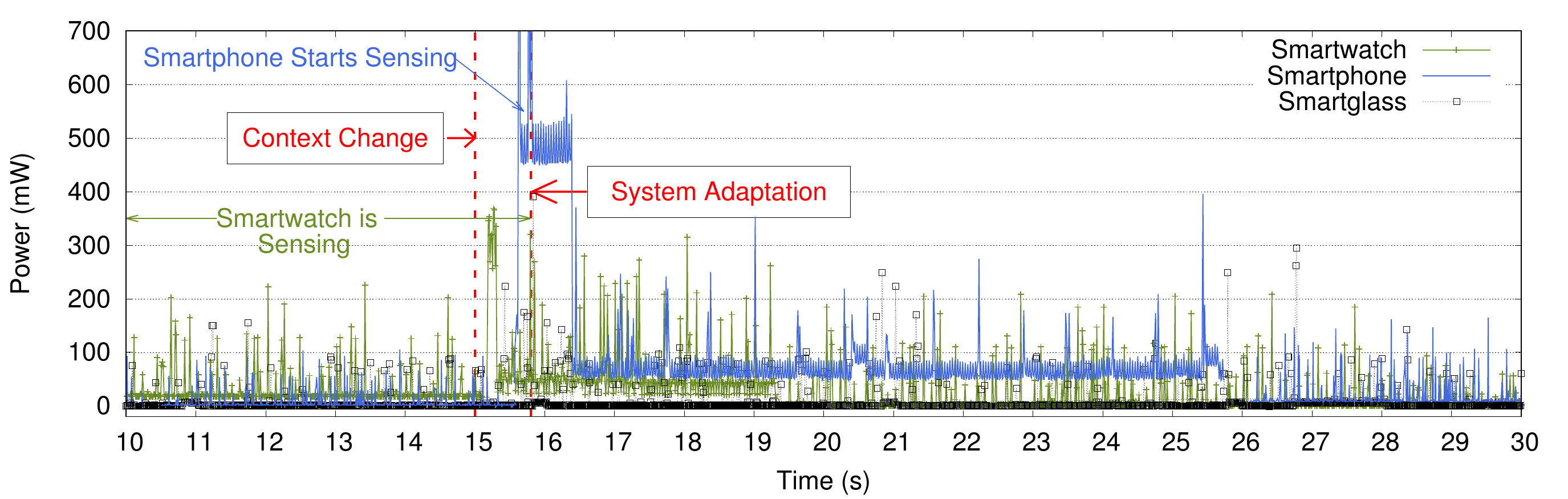}}
	\subfigure[Battery state of charge measured in smartphone and smartwatch during \sdk-enabled and not enabled cases.]{\label{fig:adapt_SOC}\includegraphics[width=0.32\textwidth]{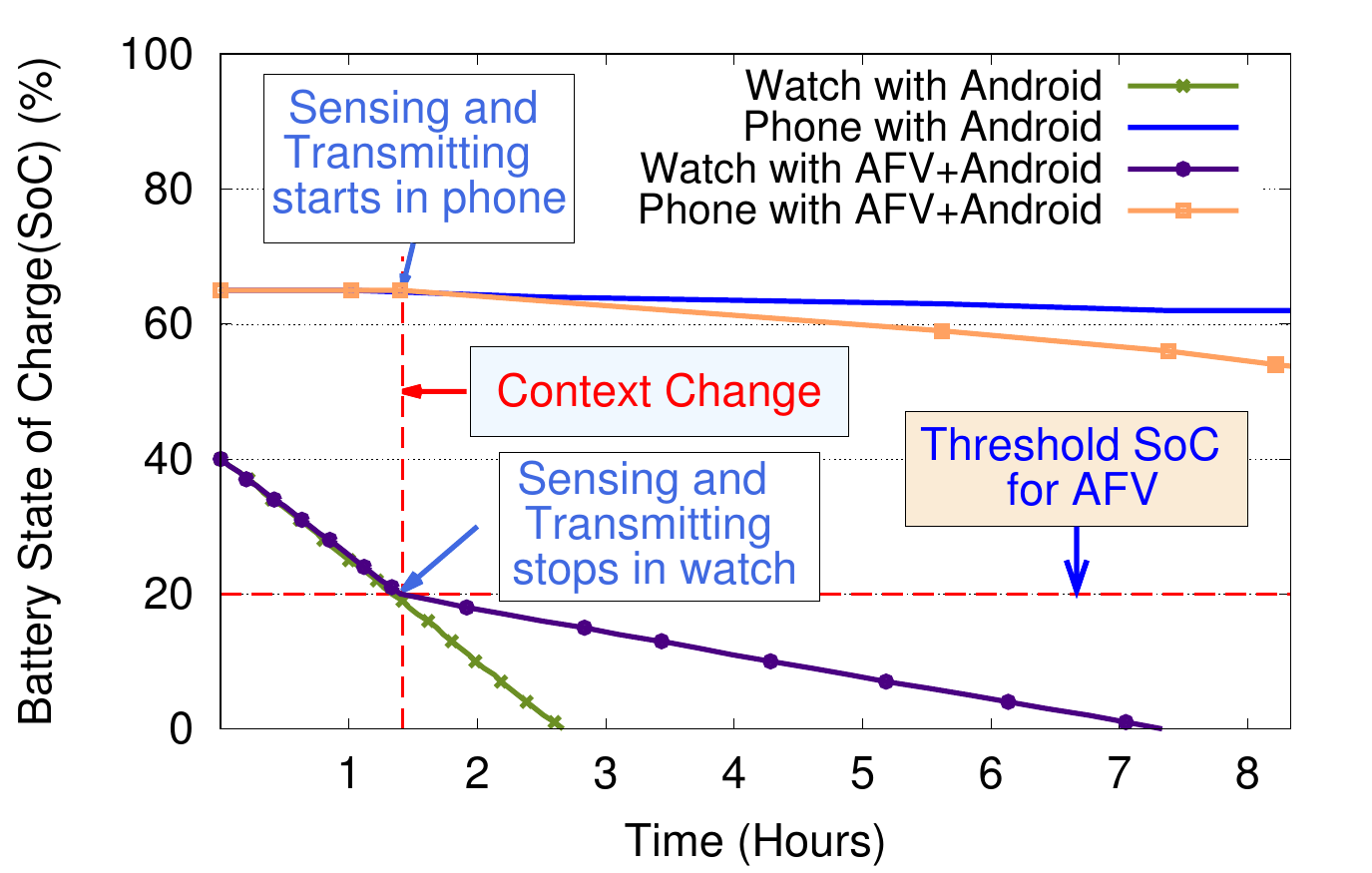}}
	\vspace {-5mm} \caption{Extending the device uptime with \sdk.}
	\label{fig:lifetime} 
	\vspace {-3mm}
\end{figure*}

\noindent\textbf{\textit{(a) Requesting accelerometer data:}} The \sdk-enabled app requires accelerometer data for fitness tracking. The user is wearing the smartglass and the smartwatch, and has the smartphone nearby while standing and exercising. At first, the user is doing head stretching exerecises and then moves to body stretching exercises. After a while, the user starts walking, carrying the smartphone in the pocket. We consider that the smartphone provides the best quality information when user is walking, smartwatch provides the best quality information when user is doing body stretching activities, and the smartglass provides the best quality information when doing head stretching exercises. This rule is used in the \emph{Decision Engine} in order to feed the app with best quality of data.

At first, while the user is doing head stretching exercises, as there is no walking detected by the smartphone on the table, and no activities are detected by the smartwatch, the smartglass performs the accelerator function and feeds to the \sdk-enabled app. When the user starts doing body stretching exercises, the accelerometer sensing function is moved to the smartwatch and feeds data to the \sdk-enabled app. When the smartphone detects that the user is walking, the sensing function moves from the smartwatch to the smartphone. Figure~\ref{fig:new_usecase1} shows the accelerometer data that is received by the \sdk-enabled application installed in the smartphone. In order to avoid unnecessary functionality movements between devices for this context, \sdk~triggers a function placement change only if the new activity continues for a goven period of time, in this experiment 10 seconds. During this time, the previously selected device continues to feed data to the \sdk-enabled app. 

\noindent\textbf{\textit{(b) Requesting heart rate data:}} Next, the \sdk-enabled app requests for heart rate (HR) data. Assume that at first, the user has the smartphone and the smartwatch. 
When the user dons the smartshirt, it is paired with the smartphone. Although the \sdk~architecture is not installed in the smartshirt (as being a Tier 2 device), it is considered as a remote sensor. 
The smartphone is responsible for pulling data from its remote sensors and feeding it to the \sdk.  

Assume that the best heart rate measurements are given by the smartshirt as it is dedicatedly designed for sensing. In the absence of the smartshirt, the data is provided by the smartwatch. 
Therefore, the \emph{Decision Engine} will select the smartshirt when ever it is available. The context change of the availability of the smartshirt via the smartphone triggers the \sdk~architecture and the \emph{Decision Engine} selects the smatrshirt  via the smartphone to feed the \sdk-enabled app. 

Figure~\ref{fig:new_usecase2} shows the heart rate data received by the \sdk-enabled smartphone app every 10 seconds. It illustrates that the smartshirt is feeding stable and accurate HR data when the user is doing the same activity. In this particular case, the smartshirt's data is available for  third party application development   
via the cloud. Therefore, the phone is connected to the cloud and retrieves the  smartshirt's real-time data and feeds it to \sdk. However, managing resources in Tier 2 devices depends on the accessibility provided by the device manufacturers. 

\noindent \textbf{(2) Maximizing the network throughput.} In this use case, the \sdk~virtualizes the Internet connectivity function in order to maximize the network throughput in a heterogeneous environment. 

Assume that an \sdk-enabled app on the smartphone requires to upload sensor data from the smartphone to an external server periodically (i.e. per second). We created two WiFi networks with different throughputs to emulate the heterogenous network. At first, the smartwatch has Internet connectivity but not the smartphone. After a while, we enable another higher speed network, to which the smartphone is connected. The device's connectivity to a new network triggers a context change which invokes the \emph{Decision Engine} to select the higher throughput network to upload the file.

At first, when the smartphone does not have direct Internet connectivity, the data from the smartphone is uplaoded to the Internet servers by relaying through the smartwatch. After a while, when direct Internet connectivity for the smartphone becomes available, the \sdk-enabled app on the smartphone automatically suspends the data transfer via the smartwatch and starts transferring directly to the Internet. 
The achieved throughput is measured at the access points by using a network analyzer (Wireshark\footnote{\url{https://www.wireshark.org/}}). Figure~\ref{fig:adapt_BW} depicts the throughput at the access points for data uploads before and after context change. 

%

\subsubsection{Extending the device's uptime.} 
We use the same \sdk-enabled app used previously. In addition, we developed two other apps for the smartphone and smartwatch that have the same functionality but do not use \sdk~(default Android). In both cases, the app installed in the smartphone requests accelerometer data, and the user preferred the app to get accelerometer data from the smartwatch which is transmitted to the smartphone once per minute. 

In the case of \sdk, when the SoC reaches the threshold (i.e. 20\%), it triggers a context change. Figure~\ref{fig:adapt} illustrates the devices' power profiles during the context change. Initially in this experiment, the smartwatch is sensing at normal speed and sending data to the smartphone once in a minute. When the context is changed at $t=15$ seconds, the smartwatch broadcasts the context change to the \emph{Master Device}, which triggers the \emph{Decision Engine} on the \emph{Master Device} to selects smartphone with a higher SoC for sensing, and informs devices.  
The smartwatch then stops sensing and the smartphone takes over the sensing function. 

The high power peaks of all devices after $t=15$ seconds is due to the messages received and transmitted by each device, which is followed by high power idle states. The high power idle state is longer for the smartphone (until $t=26$ seconds) compared to the smartwatch (until $t=20$ seconds). Figure \ref{fig:adapt} shows that the delay of system adaptation to context changes is less than one second as the smartphone starts sensing even before $t=16$ seconds. 

 
In the default case, the smartwatch keeps running its accelerometer and transfers data to the smartphone until its SoC reaches 0\%. 
Figure~\ref{fig:adapt_SOC} shows results for the increased longevity of the smartwatch battery when using \sdk. The smartwatch uptime is increased by 5 hours due to the sensing function offloading. 


\subsubsection{Minimizing the monetary cost of data usage} 

\begin{figure}[tb!]
	\centering 
	\includegraphics[width=0.45\textwidth]{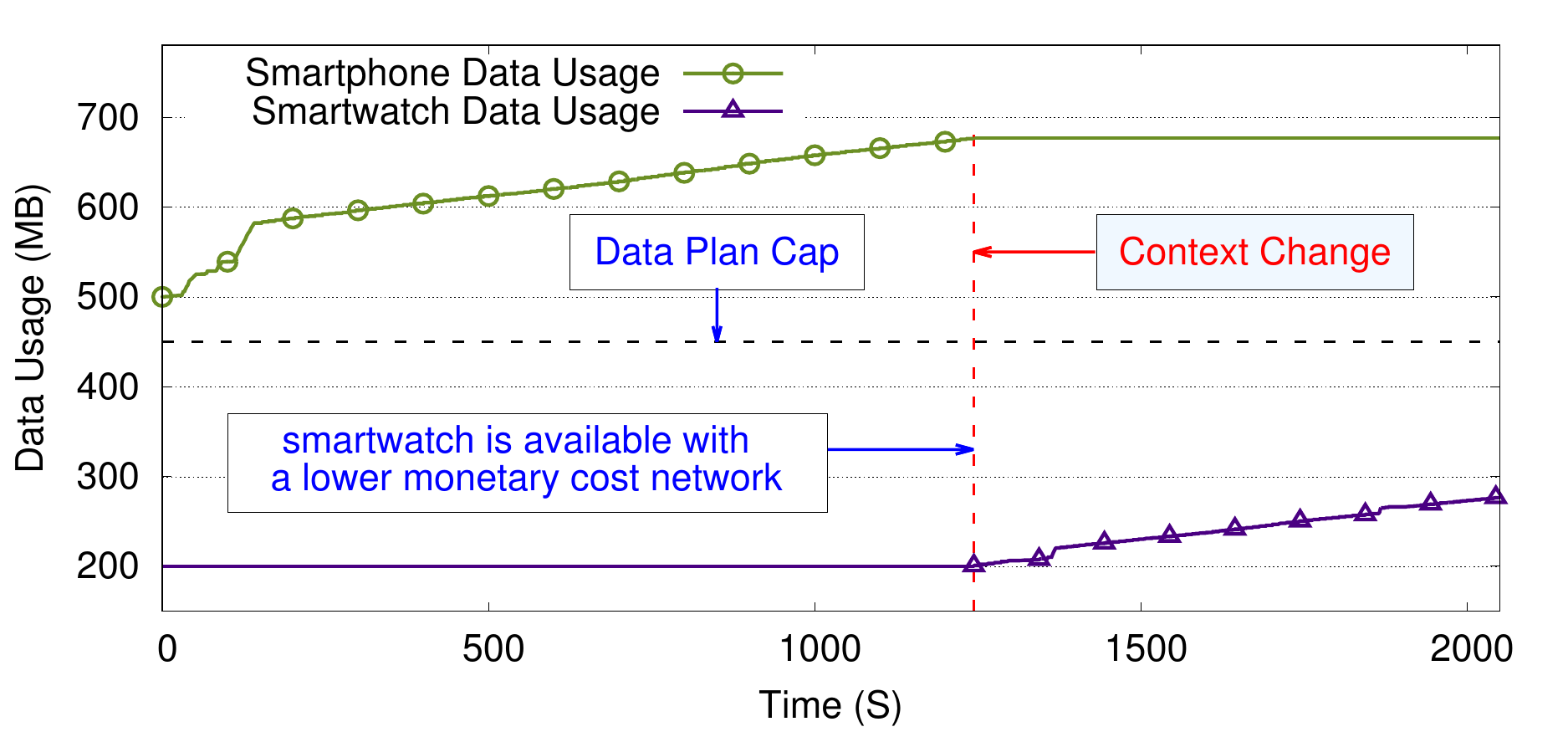}
	\vspace {-3mm}
	\caption{Dynamically switching to the less costly network for data uploading.}
	\vspace {-3mm}
	
	\label{fig:adapt_Monetary}
\end{figure}

We virtualize the Internet connectivity function in order to minimize the monetary cost of data transfer. The monetary cost for each data plan is pre-configured at system bootstrap and can be changed at any time via the \sdk~user interface. Assume that the \sdk-enabled app needs to upload files to the Internet. Also assume at first, only the smartphone has Internet connectivity, but the smartphone's data plan has exceeded the available data cap and excess data costs \$0.10/MB. After a while, the smartwatch's Internet connectivity, which has not exceeded the data cap, becomes available. Since this data plan has not exceeded the data limit, this plan costs \$0.00/MB. 

The availability of the additional network connection triggers a context change and the decision is made to use the low cost network. \sdk~notifies the device connected to network with lower cost to take over the connectivity function. 
Figure~\ref{fig:adapt_Monetary} shows the data usage of the smartphone and smartwatch before and after the context change.

\section{Related work}
\label{sec:related}



\subsection{Context monitoring}
There has been substantial work on context awareness and sensing for mobile apps. Always-on sensing can quickly drain battery resources \cite{Shen:2015}, yet continuous context monitoring is essential for proper response to context changes \cite{seemon}. This suggests a distinction between always-on and continuous that does not degrade application adaptivity nor battery life. 
These tradeoffs are explored in several research projects.

The most comprehensive sensing framework is SeeMon \cite{seemon}. Their approach leverages the relationship between sensor values and higher level ``context" states to minimize the number of sensors and their associated energy costs while continuously recognizing context changes. Another approach to sensing is to use a low-powered sensor processor to save energy. MobileHub \cite{Shen:2015} provides a framework that determined optimized alerts and submission of sensor data that reduce energy without affecting application semantics. Our context has a limited number of sensors and a small number of devices capable of performing context recognition, therefore, we have simplified the evaluation of sensor readings with a call-back mechanism for each activated sensor to inform the smart device regarding changes to the value of interest.


\subsection{Single-device resource utilization}
Adaptive system/framework for the context changes is a key concept in resources utilization.
Adaptive systems designs have been in existence for nearly 20 years \cite{randell-2005}. Early work provided context based systems development \cite{Chu2011}, prototype implementations \cite{adaptive-wearables-korteum1998, Conti2012}, programming language support for existing applications \cite{rapidware-2002}, and architectures for system design \cite{Edwards:2002, Smailagic:2002}. 
Applications on commercially available devices have only recently been deployed, due to the challenges of battery and device form, among other issues \cite{personal-wearables-2014, industry-wearables-2014, Rawassizadeh:2014}.
For the purpose of resource utilization,  Martins \cite{Martins2015} aims to tune
 the background applications in Android selectively to improve the battery lifetime. They use an OS mechanism to control the frequency of handling background tasks. 

CAreDroid~\cite{caredroid} is a framework in which to design Android applications to select the most appropriate functions to run for a given application on a single device depending on the context. It takes care of context-monitoring, adaptation decisions and allows the developer to focus on application logic only. Their work provided the inspiration for our focus on distributed system applications for wearable computer network applications. \sdk~differs in that it provides seamless function placement across devices of a PAN and function sharing across applications. Our optimization engine runs as a lightweight separate process on Tier 1 devices and the adaptation selects which functions from which devices are active at any point in time and what communication strategy will be deployed. 

Senergy \cite{senergy:2013} utilizes the sensing functionality in a way that it reduces the energy usage. This work does not require programmer intervention via the Latency, Accuracy Battery (LAB) abstraction. These 3 components are the main considerations in our framework as well, as they provide a meaningful set of tradeoffs for the user and the developer. The authors develop classifiers to infer context in sensing applications, while we use a simpler sensing strategy, but provide adaptation to achieve application goals. 

All of the above specified work consider a single device for the resources utilization. In contrast to these work, \sdk~targets to utilize the resources in a network of multiple smart-wearable devices in order to achieve the user selected objective at the runtime.

\subsection{Multi-device resource utilization}

Mechanisms to utilize the resources from multiple devices has been receiving increasing attention, and the adaptive framework designs enable developers to create tasks for multiple devices \cite{Kaler:2010, Ravindranath:2012}. There are several studies such as ErdOS \cite{Vallina-Rodriguez:2011}, CoSense \cite{Hemminki2013}, OSone \cite{osone-pasztor2013} and M+ \cite{MobilePlus}.

ErdOS leverages resources in nearby devices based on user modeling and stated user preferences. It uses a lightweight IPC and network stack to securely broadcast important context information and application data in a user-level communication manager. 
The implementation in CoSense distributes the sensing tasks between familiar devices that are in close proximity. The group formation is done by the cloud backend once the devices are registered to the cloud backend with their mac address. Once the groups are formed, the data is transferred via local connection.   

OSone distributes the functionality of the operating system in a similar fashion to how Barrelfish \cite{Baumann-barrelfish} separates functionality onto different cores. The architecture consists of a kernel node in charge of various host nodes that can be kept simpler. 
M+ allows cross-device functionality sharing. It uses remote procedure call scheme based on the binder IPC mechanism to utilize application and system functionalities across devices.

Moreover, the work such as in Reptor \cite{Reptor} allows third parties to easily distribute their modifications for a platform without the need to update the entire platform. This provide ease for the open innovation for the multi-device platforms for resources utilization. In contrast to the above systems, Rio \cite{Sani2014} presents a system that a device's resources are utilized by remotely accessing them with the help of another device in the close proximity.

We implement \sdk~in a similar fashion with the potential to have multiple controlling nodes over time, depending on remaining resources and application needs. \sdk~runs on a PAN where the devices are already connected with each other via Bluetooth. Therefore, dynamic group formation is not considered in \sdk. Moreover, most of these systems are designed for the optimal usage of sensing function. However, \sdk~framework consider all the available common functions (i.e., sensing, connectivity and computing) in the wearable personal area network. Also, we consider different optimization objectives than energy related matics such as network quality, the quality of the functionality and also the monetary cost.

We follow the philosophies initiated in the early design work on wearables. In particular, Speakeasy \cite{Edwards:2002} motivates the need for domain independent interfaces, mobile code, and user interpretation of semantics. Our representation of context is similar to that provided by Speakeasy and we retain user discernment as well. We are less ambitious in the overall goals as we leverage existing APIs and focus on the adaptive nature of wearable network applications. We do not implement code migration; we provide a system-level extension to code already available on wearable devices. Smailagic and Sieworek \cite{Smailagic:2002} provide key design principles/challenges for future wearable applications: user interface models, input/output modalities, matched capability with requirements, and quick interface evaluation methodology. We focus on the third of these challenges to meet the user's needs with the lowest resource utilization.


\section{Discussion}

\sdk~ is designed with the premise that all devices in a PAN are always connected and managed by the same person. Therefore, we have not considered the option of dynamic device group formation with nearby devices owned by other people. This is primarily to reduce the privacy and security concerns of communicating with untrusted devices of strangers. On the other hand, we assumed that there are no privacy or security risk in communicating or utilizing functions on the trusted devices on the same PAN. However, this assumption may not always be true, as the third-parties such as trackers, intruders and manufacturers have the access and partially control some functionalities of the devices and its data. Therefore, we intend to mitigate the potential threats of information leakage with the PAN by extending \sdk~ with a context-aware security framework incorporating a set of pre-defined and also user defined device access policies. These policies will then be considered during function allocation as another context information. For example, if a device is connected to a public WiFi access point, the device may not be used to virtualize functions by the \emph{Decision Engine}.

In this paper, we have only validated \sdk~performance for limited functionalities, i.e. sensing and Internet connectivity, although \sdk~ is designed for efficient utilization of many other functions such as compression, encoding and anonymization. We have noticed the potential difference in overheads associated with each function. Therefore, we aim to further strengthen our function allocation algorithm considering available memory and CPU power as additional context. In addition, for each particular use case, we considered a single objective, which are specified in Table \ref{tab:context}. However, user may wish to achieve multiple objectives at the same time. As an example, user may wish to have the best quality of information while achieving the minimum possible energy consumption. This can be addressed by formulating a mutli-objective optimization problem in the \emph{Decision Engine} with different weighting factors for each of the objective. These weighting factors are to be specified by the user via the UI provided by \sdk.  

Although we developed \sdk~ prototype as a standalone user-level application and a library, it can also be realized by integrating to the OS as a module, which requires root access permission to the kernel. OS module implementation will be efficient in terms of systems overheads of \sdk, but it requires significant development effort as well as reduces the deployability of \sdk. However, despite this implementation overheads, we showed that \sdk~outperforms the vanilla scenario without \sdk. Therefore, we aim to further improve the user-level development to release \sdk~as a software development kit (SDK) for app developers, and also, envisage the implementation in multiple OSs.


\section{Conclusion}
Majority of devices in a personal area network that consists of multiple smart wearables and hand-held devices have a number of common capabilities and resources. However, current popular mobile and wearable applications do not utilize these resources efficiently that leads to multiple of application function executions in the same personal area network. As a result, the users may not get the best outcome, may incur higher networking cost and may also result in higher energy consumption of devices. 


In this paper, we proposed \sdk,  an architecture that overcomes the above inefficiencies, while reducing the overall energy usage, without adding any latency and minimizing the communication overhead. \sdk~enables context-aware application function virtualization in a personal area network with a set of APIs that can be easily leveraged by app developers during application development. Our simulation results showed that the proposed function allocation algorithm enables system uptime improvement of up to 35-40\% compared with typical configurations of current wearable/mobile applications.
Then, we showed the viability of the architecture by implementing \sdk~in Android devices without loss of generality. Finally, we showed the real world applicability of \sdk~and user benefits via emulating multiple use cases with real devices.

\appendices
\section{}

The defined message formats for Inter-device communication are shown in Table~\ref{tab:dataStructure}. These messages are byte streams and start with an eight bit ID field that is reserved to announce the \texttt{message type}. 

\begin{table}[tb]
	\caption{Message formats.}
	\centering
	\scriptsize
	\begin{tabular}{|l|c|}
		\hline
		
		\textbf{Description}& \textbf{Length [Bytes]} \\
		\hline
		\hline
		\multicolumn{2}{|l|}{\textbf{Initialization Message}}\\\hline
		Message Type  & 1\\\hline
		Device ID & 8\\\hline
		Device type (Phone/Watch/Glass)  & 1\\\hline
		Number of networks in the array & 1\\\hline
		Length of ID, & \multirow{2}{*}{$<$ 1, $n_1$, 4 $>$}\\
		SSID/ Operator ID, Monetary cost & \\\hline
		Number of function types in the array & 1\\\hline
		Function type, Energy (per function) & $<$ 1, 4 $>$\\\hline
		\hline

		\multicolumn{2}{|l|}{\textbf{Context (Sensor) Message}} \\
		\hline
		Message Type  & 1\\\hline
		Battery level & 1\\\hline
		Charging status & 1\\\hline
		Moving status & 1\\\hline
		Connected network type (Wifi/Cellular), &  \multirow{2}{*}{$<$ 1, 1, $n_2$ $>$}\\
		Length of ID, SSID/ Operator ID& \\\hline
		
		Average link speed & 4\\\hline
		\hline
		
		\multicolumn{2}{|l|}{\textbf{Context (Request) Message}}\\\hline
		Message Type  & 1\\\hline
		Request type  & 1\\\hline
		Request related information  & $n_3$\\\hline
		\hline
		
		\multicolumn{2}{|l|}{\textbf{Assignments Message}}\\\hline
		Message Type  & 1\\\hline
		Length of the $R_d \mapsto D$ mapping  & 1\\\hline
		$R_d \mapsto D$ $<$Request, Device$>$  & $<$1,1$>$\\\hline
		Length of the $V_d \mapsto D$ mapping & 1\\\hline
		$V_d \mapsto D$ $<$Function, Device$>$  & $<$1,1$>$\\\hline
		\hline
		
		\multicolumn{2}{|l|}{\textbf{Data Message}}\\\hline
		Message Type  & 1\\\hline
		$<$Request type, Data length,Data$>$  & $<$1,4,$n_4$$>$\\\hline
		
	\end{tabular}
	\label{tab:dataStructure}
\end{table}

\textbf{Initialization Message:} The message starts with the 1) \texttt{device ID} and \texttt{device type}, a combination that is used to uniquely identify the device in the PAN followed by connected/available network information and supported functions information. 

\textbf{Context (Sensor) Message} contains the source \texttt{device ID} along with the updated information. The current \sdk~prototype implementation is limited to the context information detailed in Table~\ref{tab:dataStructure}.  

\textbf{Context (Request) Message} contains an identifier for the \texttt{request type} (e.g., accelerometer data) and the request related information. The \texttt{request related information} is unique to each request type. 
For example, sensing request contains sensing frequency, and downloading request contains a URL. Therefore, we introduce a \texttt{message length} field that can be used to parse the message byte by byte in combination with \texttt{request type}.

\textbf{Assignments Message} contains the mapping data of $R_d \mapsto D$ and $V_d \mapsto D$. 

\textbf{Data Message} transfers application functions' input/output data and contains the \texttt{data} itself along with the \texttt{request type}. For instance, \texttt{data} would be a file to upload if the request type is internet access, or would be an array of sensing information if the request type is sensing. 
\ifCLASSOPTIONcaptionsoff
  \newpage
\fi

\balance{}
\bibliographystyle{IEEEtran}
\bibliography{biblio}

%
%
%
%

\balance{}
\end{document}